\documentclass[useAMS,usenatbib]{mnras}
\usepackage{amssymb}
\usepackage{graphicx}
\DeclareGraphicsExtensions{.jpg,.pdf,.png,.eps,.ps}
\usepackage{url}
\usepackage{color}
\usepackage{amsmath}
\newcommand{\hstlong}{\textit{Hubble Space Telescope}}
\newcommand{\integral}{\textit{INTEGRAL}}
\newcommand{\swift}{\textit{Swift}}

\newcommand{\xmmlong}{\textit{XMM-Newton}}
\newcommand{\xmm}{\textit{XMM}}
\newcommand{\rosat}{\textit{ROSAT}}
\newcommand{\nicer}{\textit{NICER}}
\newcommand{\fermi}{\textit{Fermi}}
\newcommand{\nustar}{\textit{NuSTAR}}

\newcommand{\nh}{\mbox{$N_{\rm H}$}}

\newcommand{\kteff}{\mbox{$kT_{\rm eff}$}}

\newcommand{\rns}{\mbox{$R_{\rm NS}$}}
\newcommand{\mns}{\mbox{$M_{\rm NS}$}}

\newcommand{\nht}{\mbox{$N_{\rm H,20}$}}
\newcommand{\chisq}{\mbox{$\chi^2$}}
\newcommand{\chisqnu}{\mbox{$\chi^2_\nu$}}

\newcommand{\Chisq}[3]{$\chi^2_\nu$/dof (prob.) = {#1}/{#2} (#3)}

\newcommand{\xray}{\mbox{X-ray}}

\newcommand{\simlt}{\mathrel{\hbox{\rlap{\hbox{\lower4pt\hbox{$\sim$}}}\hbox{$<$}}}}
\newcommand{\simgt}{\mathrel{\hbox{\rlap{\hbox{\lower4pt\hbox{$\sim$}}}\hbox{$>$}}}}

\newcommand{\approxlt}{\mbox{$\,^{<}\hspace{-0.24cm}_{\sim}\,$}}
\newcommand{\ee}[1]{\mbox{$10^{#1}$}}
\newcommand{\tee}[1]{\mbox{$\times 10^{#1}$}}
\newcommand{\ud}[2]{\mbox{$^{+ #1}_{- #2}$}}
\newcommand{\ppm}{\mbox{$\pm$}}

\newcommand{\unit}[1]{\mbox{$\rm\,#1$}}
\def\deg{\hbox{$^\circ$}}

\def\arcmin{\hbox{$^\prime$}}
\def\arcsec{\hbox{$^{\prime\prime}$}}

\newcommand{\K}{\mbox{$\,K$}}
\newcommand{\msun}{\mbox{$\,M_\odot$}}

\newcommand{\km}{\hbox{$\,{\rm km}$}}

\newcommand{\cm}{\mbox{$\,{\rm cm}$}}

\newcommand{\keV}{\mbox{$\,{\rm keV}$}}
\newcommand{\eV}{\mbox{$\,{\rm eV}$}}
\newcommand{\ksec}{\mbox{$\,{\rm ks}$}}
\newcommand{\msec}{\mbox{$\,{\rm ms}$}}
\newcommand{\yr}{\mbox{$\,{\rm yr}$}}

\newcommand{\persec}{\mbox{$\,{\rm s^{-1}}$}}

\newcommand{\percmsq}{\mbox{$\,{\rm cm^{-2}}$}}

\def\apjl{ApJL}
\def\apj{ApJ}
\def\mnras{M.N.R.A.S.}
\def\aap{A\&A}
\def\nat{Nat.}

\def\apjs{ApJ Supp.}

\def\physrep{Phys. Rep.}

\def\prd{Ph.Rev.D}

\title{The \nustar\ view of the non-thermal emission from
  PSR~J0437--4715}

\author[S. Guillot et al.]{S. Guillot$^{1,2}$ \thanks{email:
    sguillot@astro.puc.cl}, V. M. Kaspi$^2$, R. F. Archibald$^2$,
  M. Bachetti$^3$, C. Flynn$^4$, \newauthor F. Jankowski$^4$,
  M. Bailes$^4$, S. Boggs$^5$, F. E. Christensen$^6$,
  W. W. Craig$^{6,7}$, \newauthor C. A. Hailey$^8$,
  F. A. Harrison$^9$, D. Stern$^{10}$, W. W. Zhang$^{11}$ \\ $^1$
  Instituto de Astrof\'{i}sica, Facultad de F\'{i}sica, Pontificia
  Universidad Cat\'{o}lica de Chile, Av. Vicu\~{n}a Mackenna 4860,
  \\782-0436 Macul, Santiago, Chile \\ $^2$ Department of Physics and
  McGill Space Institute, McGill University, 3600 rue University
  Montr\'{e}al, QC, Canada H3A-2T8 \\ $^3$ Osservatorio Astronomico di
  Cagliari, via della Scienza 5, 09047 Selargius, Italy\\ $^4$ Centre
  for Astrophysics and Supercomputing and ARC Centre for All-Sky
  Astrophysics (CAASTRO),\\ Swinburne University of Technology, Post
  Office Box 218 Hawthorn, VIC 3122, Australia.\\ $^5$ Space Sciences
  Laboratory, University of California, Berkeley, CA 94720, USA
  \\ $^6$ DTU Space, National Space Institute, Technical University of
  Denmark, Elektrovej 327, DK-2800 Lyngby, Denmark \\ $^7$ Lawrence
  Livermore National Laboratory, Livermore, CA 94550, USA \\ $^8$
  Columbia Astrophysics Laboratory, Columbia University, New York, NY
  10027, USA \\ $^9$ Space Radiation Laboratory, California Institute
  of Technology, 1200 E California Blvd, MC 249-17, Pasadena, CA
  91125, USA \\ $^{10}$ Jet Propulsion Laboratory, California
  Institute of Technology, Pasadena, CA 91109, USA \\ $^{11}$ NASA
  Goddard Space Flight Center, Astrophysics Science Division, Code
  661, Greenbelt, MD 20771, USA}

\begin{document}

%\date{\today}

%\pagerange{\pageref{firstpage}--\pageref{lastpage}} \pubyear{}

\maketitle

\label{firstpage}

\begin{abstract}

We present a hard \xray\ \nustar\ observation of PSR~J0437--4715, the
nearest millisecond pulsar.  The known pulsations at the apparent
pulse period $\sim5.76\msec$ are observed with a significance of
3.7$\sigma$, at energies up to 20\keV\ above which the
\nustar\ background dominates.  We measure a photon index $\Gamma=
1.50\pm0.25$ (90\% confidence) for the power law fit to the
non-thermal emission.  It had been shown that spectral models with two
or three thermal components fit the \xmmlong\ spectrum of
PSR~J0437--4715, depending on the slope of the power-law component,
and the amount of absorption of soft \xray s.  The new constraint on
the high-energy emission provided by \nustar\ removes ambiguities
regarding the thermal components of the emission below $3\keV$.  We
performed a simultaneous spectral analysis of the \xmmlong\ and
\nustar\ data to confirm that three thermal components and a power law
are required to fit the 0.3--20\keV\ emission of PSR~J0437--4715.
Adding a \rosat -PSPC spectrum further confirmed this result and
allowed us to better constrain the temperatures of the three thermal
components.  A phase-resolved analysis of the \nustar\ data revealed
no significant change in the photon index of the high-energy emission.
This \nustar\ observation provides further impetus for future
observations with the \nicer\ mission ({\it Neutron Star Interior
  Composition Explorer}) whose sensitivity will provide much stricter
constraints on the equation of state of nuclear matter by combining
model fits to the pulsar's phase-folded lightcurve with the pulsar's
well-defined mass and distance from radio timing observations.

\end{abstract}


%%%%%%%%%%%%%%%%%%%%%%%%%%%%%%%%%%%%%%%%%%%%%%%%%

\section{Introduction}

PSR~J0437--4715 (J0437, hereafter), currently the nearest millisecond
pulsar (MSP) to the Earth, was discovered in the 70\cm\ Parkes radio
pulsar survey \citep{johnston93} and subsequently in the X-rays with
\rosat\ \citep{becker93}. Since then, it has been extensively studied
in multiple energy bands.  Radio observations allow precise
measurements of the pulsar spin ($P=5.76\msec$), intrinsic spin-down
rate ($\dot{P}=5.8\tee{-20}\unit{s\persec}$), binary system orbital
properties \citep{verbiest08}, as well as the distance
$156.3\pm1.3\unit{pc}$ \citep{deller08}.  In the infrared and optical
bands, the emission from the white dwarf companion dominates, leading
to constraints on its properties \citep{durant12}.  J0437 has also
been detected in the near and far ultraviolet (UV), where the
Rayleigh-Jeans tail of the pulsar surface thermal emission is thought
to dominate the thermal emission of the white dwarf companion
\citep{durant12}.  Previous \xmmlong\ fast-timing mode observations of
J0437 have permitted the interpretation of the phase-modulated soft
\xray\ flux as the thermal emission from two polar caps moving in and
out of the line of sight as the pulsar rotates \citep{bogdanov13}.
Observations with the \textit{ Fermi Gamma-Ray Space Telescope} have
also detected $>100\unit{MeV}$ pulsed emission from J0437, thought to
originate far from the neutron star (NS) surface \citep{abdo09}.

However, because of its faintness in the hard \xray\ band ($\sim$
5--100\keV), as well as its proximity on the sky ($\sim4\arcmin$) to
the bright active galactic nucleus RX~J0437.4-4711 \citep{halpern96b},
the emission of J0437 in this band has been inaccessible to the
previous generation of hard \xray\ telescopes, such as
\integral\ \citep{winkler03} or the \swift\ Burst Alert Telescope
\citep{krimm13}.  The MOS detectors of \xmmlong\ cover part of the
hard \xray\ band ($\lesssim 10\keV$), but the drop in effective area
above 4\keV\ limits the constraints on the high energy tail of the
spectrum \citep{bogdanov13}.  When it was launched, the {\it Nuclear
  Spectroscopic Telescope Array} (\nustar), with its hard
\xray\ focusing optics capabilities, opened a new window on the hard
\xray\ sky \citep{harrison13}, with unrivalled sensitivity in the
3--79\keV\ range. In addition, the angular resolution of \nustar\ (18
arcseconds full-width half-maximum) allows us to obtain the
high-energy spectrum of J0437 without the contaminating photons from
the nearby active galactic nucleus.

With the \xmm\ observations, phase-averaged (with \xmm -mos) and pulse
phase-resolved (with \xmm -pn in fast-timing mode) analyses were used
to determine the system geometry and to characterize the surface
emission (\citealt{bogdanov13}, using the method presented in
\citealt{bogdanov08}).  In turn, these analyses led to some
constraints on the compactness of the pulsar $\mns/\rns$. The
independently measured mass, $\mns=1.76\pm0.20\msun$ from radio timing
\citep{verbiest08}, therefore restricted the range of possible radii
to values $\gtrsim 11.1\km$.  Recent radio timing measurements of
J0437 have led to a mass estimate of $\mns=1.44\pm0.07\msun$
\citep{reardon16}, which would allow slightly lower radii according to
the analysis of \cite{bogdanov13}.

Measurements of the NS radii are crucial to determine their
interior properties, and study the behavior of dense nuclear matter
\citep[e.g.,][for reviews]{lattimer07,heinke13,miller13}.
Specifically, the equation of state of matter at and above nuclear
density can only be determined by observations of NSs, and
in particular, measurements of their radii.  In addition to the pulse
profile modeling of millisecond pulsars, other methods exist to
measure the NS radius, including constraints from the
photospheric radius expansion bursts of type I \xray\ bursting NSs
\citep[e.g.,][]{ozel09b,guver10a,suleimanov11b,ozel16}, and the
measurement of \rns\ from the surface emission of NSs in quiescent
low-mass \xray\ binaries
\citep{webb07,heinke06a,heinke14,guillot11b,guillot13,guillot14}.
However, all methods are affected by systematic uncertainties and it
is important to understand and test them all until they converge to
consistent results.

For J0437, the measurement of the compactness (i.e., radius
measurement) relies on observations of the pulsed thermal surface
emission.  However, it has been shown that the poorly constrained
high-energy tail, modeled with a power law (PL), affects how well the
thermal emission can be constrained \citep{bogdanov13}. Specifically,
depending on the PL index, two or three thermal components were
necessary to fit the soft X-ray emission of J0437: with a softer PL
index $\sim$ 2.5, the third thermal component ($\kteff\sim50\keV$) is
unnecessary.  This can be explained by the fact that a PL with
$\Gamma\sim2.5$ fitted to the high-energy tail starts dominating the
emission in the 0.1--0.3\keV\ range where the fluxes from the two
hottest thermal components (corresponding to the spot emission)
decline.  On the other hand, a harder PL with $\Gamma\sim1.5$ leaves
an excess of counts in the 0.1--0.3\keV\ band which is then well fit
by a cold thermal component.

Since the constraints on $\mns/\rns$ are extracted from the pulse
profile and the thermal components normalizations, the poorly
constrained PL limits the precision of the $\mns/\rns$ measurement.
In other words, an independent measurement of the PL index can provide
a better constraint on the thermal components and therefore on the
radius of J0437.  This is of particular importance since this MSP is
the primary target for the upcoming \emph{Neutron Star Interior
  Composition Explorer} (\nicer, \citealt{gendreau12}, expected launch
date: early 2017).  The large effective area of this future mission
mounted on the International Space Station will collect $\sim\ee{6}$
photons from J0437 (amongst other NSs) for the sole purpose of
analyzing its pulsed emission and measuring its compactness.  It is
therefore crucial to know and constrain, ahead of time, the nature of
the high energy emission of this key \nicer\ target.

This paper presents a \nustar\ observation of J0437, as well as a
joint analysis with archival \xmm\ observations.  Analysis of an
archival \rosat-PSPC observation complements the \xmm-MOS data by
providing an energy coverage to lower energies.  Combined with the
\nustar\ observation, the spectral data presented in this work cover
an energy range from 0.1\keV\ to 20\keV. In Section~\ref{sec:obs}, the
observations are described together with the data reduction and
analysis procedures.  The results of the timing and spectral analysis
are presented in Sections~\ref{sec:timing} and
\ref{sec:spectralAnalysis}, respectively.  Finally,
Section~\ref{sec:ccl} summarizes and discusses the results.

\begin{table}
 \centering
 \caption{Selected \xray\ Observations of PSR~J0437--4715}
 \resizebox{\columnwidth}{!}{%
   \begin{tabular}{cccc}
     \hline
      ObsID & Telescope/  & Start     &  Usable Exposure \\
            & Instrument  & Time (UT) & (ksec)  \\
     \hline
      701184      & \rosat-PSPC  & 1992-09-20 17:51:32 &   6.1 \\
      0603460101  & \xmm-mos1    & 2009-12-15 19:41:16 & 118.2 \\
      0603460101  & \xmm-mos2    & 2009-12-15 19:41:16 & 119.3 \\
      0603460101  & \xmm-pn      & 2009-12-15 20:00:24 & 117.0 \\
      30001061002 & \nustar-FPMA & 2014-12-29 14:51:07 &  78.8 \\
      30001061002 & \nustar-FPMB & 2014-12-29 14:51:07 &  78.8 \\
      30001061004 & \nustar-FPMA & 2014-12-31 18:26:07 &  68.4 \\
      30001061004 & \nustar-FPMB & 2014-12-31 18:26:07 &  68.4 \\
      30001061006 & \nustar-FPMA & 2015-01-02 17:16:07 &  67.9 \\
      30001061006 & \nustar-FPMB & 2015-01-02 17:16:07 &  67.9 \\
     \hline
     \label{tab:Obs}
   \end{tabular}   }
\end{table}

\section{Observations and Analysis}
\label{sec:obs}

\subsection{\nustar\ observations}
The \nustar\ Observatory consists of two co-aligned telescopes
focusing hard \xray s in the 3--79\keV\ range onto two focal planes
modules, FPMA and FPMB \citep{harrison13}.  It provides relatively
low-background imaging capabilities (18\arcsec\ full-width
half-maximum) in the hard \xray\ band with a $2\unit{\mu sec}$
relative timing resolution.

A $\sim 200\ksec$ (total on source) observation of J0437 was performed
in Dec./Jan. 2015 (see Table~\ref{tab:Obs}).  The FPMA and FPMB data
were re-processed following the standard pipeline of the \nustar\ data
analysis system, \emph{nupipeline} v0.4.3 and \emph{heasoft} v16.0,
together with the \nustar\ calibration files from \emph{CALDB} v4.6.8.
During the data processing, photon time tags were barycentered to the
Solar system reference frame using the position of J0437 obtained from
radio timing.

For the spectral analysis, the size of the circular extraction region
was chosen to maximize the signal-to-noise ratio (S/N).  Specifically,
a range of source extraction radii were tested between 10\arcsec\ and
120\arcsec , with various annuli around the source regions for the
backgrounds.  A 35\arcsec\ circular source region maximizes the S/N
and was therefore used to generate the spectra. For the background, an
annulus region with inner radius 80\arcsec\ and outer radius of
87.2\arcsec\ was used to ensure that both the source and background
regions have the same area.  Because the background noise dominates
the source signal above 20\keV, counts above that energy are
discarded, restricting the spectral analysis to the 3--20\keV\ energy
range.  The background subtracted count rates for J0437 are
$1.50\pm0.15\unit{counts/ksec}$ for FPMA and
$1.60\pm0.16\unit{counts/ksec}$ for FPMB, in the 3--20\keV\ energy
range.  Data from each module were combined to generate one spectrum
for each module (including the response files), and following that
step, photon events were grouped with a minimum of 40 counts per bin.
Note that the FPMA and FPMB spectra are not combined together.
Instead a cross-normalization constant between them is used during the
spectral analysis to account for cross-calibration uncertainties
between the two modules (see Section~\ref{sec:nustar}).

\subsection{\xmmlong\ observations}
This work used a continuous 130\ksec\ archival observation of J0437
whose analysis was presented in detail in a previous work
\citep{bogdanov13}.  The fast readout cadence (30\unit{\mu sec}) of
the European Photon Imaging Camera (EPIC) pn instrument in timing mode
allowed the measurement of the pulse profile of J0437, while the EPIC
MOS1 and MOS2 instruments were used for the spectral analysis.  In
this work, the raw pn and MOS data were reprocessed with the
\emph{pnchain} and \emph{emchain} tasks of the \emph{XMMSAS} v14.0
package and with the latest calibration files, following the standard
data reduction and analyses procedures.  As in \cite{bogdanov13}, the
time intervals of high background were removed leading to the exposure
times listed in Table~\ref{tab:Obs}.  For the MOS1/2 imaging data,
circular extraction regions were used with 60\arcsec\ radii for the
source regions.  The MOS1/2 spectra were generated in the
0.3--10\keV\ band since the calibration below 0.3\keV\ is unreliable.
Specifically, quantum efficiency inhomogeneities in the detector
itself are present at low energy (see \xmm\ calibration document
CAL-TN-0018\footnote{Available at
  \url{http://www.cosmos.esa.int/web/xmm-newton/calibration-documentation}}), and are
not considered in the \emph{XMMSAS} (priv. comm., J. Ebrero,
\xmmlong\ Science Operation Center).  For all spectra, the response
matrix and ancillary response files were obtained with the tools
\emph{rmfgen} and \emph{arfgen}, respectively.  The MOS1 and MOS2
spectra (extracted with identical parameters) and their response files
were combined into a single spectrum using the task
\emph{epicspeccombine}, and events were then binned with a minimum of
100 counts per bin.  Overall, the data reduction of the \xmm\ data was
very similar to that of \cite{bogdanov13}, except for the energy range
used.

A 2002 observation of J0437 with \xmm\ is also publicly available.
However, the \xray\ flux decreased by $\sim 5\%$ between 2002 and
2009.  This decrease was measured by fitting the spectra of both
epochs with the same model, except for a multiplicative constant
(fixed to unity for the 2002 data, and free to vary for the 2009
data).  We found a statistically acceptable fit, and a multiplicative
constant $c=0.945\pm0.014$ (90\% confidence). We note that an
accumulation of contaminants on the MOS detectors since the launch of
\xmm\ have affected the sensitivity in the low-energy
range\footnote{see \xmm\ calibration document CAL-SRN-0305 available
  at
  \url{http://xmm2.esac.esa.int/docs/documents/CAL-SRN-0305-1-0.ps.gz}}.
However, this effect has been accounted for in the publicly-available
\xmm\ calibration files since \emph{XMMSAS} v13.5.  It is beyond the
scope of this paper to present a detailed analysis of the soft
\xray\ variability of J0437. We therefore choose to use only the
longest and most recent observation of this MSP, and we leave it for
future work to investigate the origin of this possible soft
\xray\ variability.

\subsection{\rosat-PSPC observations}
This work also took advantage of the energy response of the
\rosat-PSPC camera down to 0.1\keV.  Using pre-processed data sets,
the spectral extraction was performed with the {\tt ftools} package
{\tt xselect}.  Source counts were selected with a circular region
centered around the radio position of J0437, with a radius of
70\arcsec.  For the background, a source-centered annulus was chosen,
with inner radius 70\arcsec\ and outer radius 110\arcsec.  The
response matrix file is provided by CALDB\footnote{Available at
  \url{https://heasarc.gsfc.nasa.gov/docs/rosat/pspc_matrices.html}}
and the ancillary response file with the ftools command \emph{pcarf}.
Channels were grouped with a minimum of 40 counts per bin.

\bigskip

\section{Timing Analysis}
\label{sec:timing}
\begin{figure}
  \centering
  \makebox[0cm]{\includegraphics[width=8.2cm]{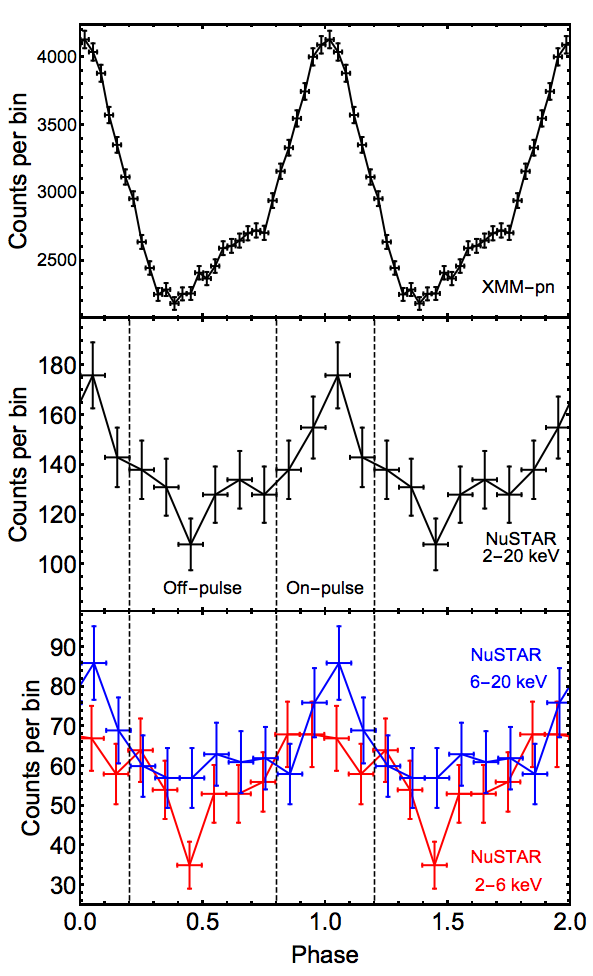}}
  \caption{ {\it (top)} XMM-pn pulse profile of PSR~J0437--4715 in the
    0.5--2.0\keV\ energy range, as analysed in
    \citet{bogdanov13}. {\it (middle)} \nustar\ pulse profile of
    PSR~J0437--4715 obtained by folding photons in the 2--20\keV\ band
    at the known radio ephemeris. The H-test probability associated
    with this pulse profile is 0.00022 ($\sim3.7\sigma$).  The
    vertical dashed lines separate the on- and off-pulse regions used
    to generate the phase-resolved spectra presented in
    Section~\ref{sec:PhaseRes}.  {\it (botton)} 2--6\keV\ and
    6--20\keV\ \nustar\ pulse profiles of PSR~J0437--4715.  When the
    photons are split into two energy bands, the pulsation detection
    significance diminishes (see text) due to the decrease in
    S/N. Note that the pulse profiles of PSR~J0437--4715 observed by
    \nustar\ might be distorted from the true morphology by the
    \nustar\ clock drift (see Section~\ref{sec:clockdrift}).}
  \label{fig:pulse}
\end{figure}

To study the 2.0--20\keV\ \nustar\ pulse profile of J0437, the
barycentered photon event times were folded using the radio ephemeris,
obtained at the Molonglo Observatory (see
Appendix~\ref{sec:appendix}), with the {\tt photon} plugin\footnote{by
  Anne Archibald, based on Lucas Guillemot's {\tt Fermi} plugin.} of
the TEMPO2 pulsar timing package \citep{edwards06}.  The phases
obtained from folding were then grouped into 10 phase bins, with
Poisson errors in each bin (Figure~\ref{fig:pulse}).  An H-test was
performed to evaluate the significance of the observed pulse
\citep{dejager89}.  We found an H-test probability of 0.00022, which
quantifies the significance of the H-value by estimating the
probability of obtaining, by chance, an H-value as large, or larger,
from sampling a uniform distribution.  The H-test probability found
corresponds to a $\sim3.7\sigma$ significance.  By separating photons
from the on- and off-pulse parts of the profile (see divisions in
Figure~\ref{fig:pulse}), spectra were obtained during these two
different phases of the rotation.  This is further discussed in
Section~\ref{sec:PhaseRes}.

The resulting pulsed fraction, defined as the fraction of counts above
the minimum\footnote{This pulsed fraction estimator is used here for
  convenient comparison with the work of \citealt{bogdanov13}.  Note
  that there is a possible bias toward higher pulse fraction with this
  estimator compared to the root-mean-square (RMS) pulsed fraction
  estimator \citep{an15}.  Nonetheless, for a small number of bins
  ($\sim$10 bins, as used in the present work), the bias compared to
  the RMS method appears to be minimal (see Fig.~12 of
  \citealt{an15}).}, of the J0437 profile is 24\ppm 6\% ($1\sigma$,
all pulse fractions errors provided thereafter are given at the
$1\sigma$ level), for photons in the energy range 2--20\keV.  Note
that \nustar\ is sensitive to photons below $3\keV$, and while those
photons cannot formally be used for a spectral analysis (due to the
current absence of calibration below $3\keV$), they were used here to
study the pulse profile.  When compared to the pulse fractions
measured in the soft \xray\ band with \xmm\ (32\ppm1\%, 35\ppm1\%,
37\ppm1\%, 37\ppm1\%, and 35\ppm2\% in the 0.275-–0.35, 0.35-–0.55,
0.55-–0.75, 0.75-–1.1, and 1.1-–1.7\keV\ bands, respectively,
\citealt{bogdanov13}), there appears to be a decrease in the pulsed
fraction as energy increases.  Photons were then split into two
separate bands, 2--6\keV\ and 6--20\keV, before being folded with the
method described in the previous paragraph.  The separation energy is
chosen to have similar number of photons in both bands.  The pulse
fractions in these two bands are: 32\ppm 9\% and 20\ppm 8\%,
respectively.  These pulse fractions are different from 0 at the
99.92\% and 98.76\% confidence level, respectively, which correspond
to 3.4$\sigma$ and 2.5$\sigma$.  Although the S/N does not allow one
to draw a firm conclusion, these values appear to reinforce the
suggestion that the pulsed fraction decreases with increasing photon
energy.  We also note that although the detection of pulsations in the
full 2--20\keV\ energy range is significant, the detection is only
marginal in the two separate energy bands: the H-test probabilities
are 0.0057 for the low energy band (equivalent to 2.7$\sigma$), and
0.015 for the high energy band (equivalent to 2.4$\sigma$).

It is important to keep in mind that the true morphology of the pulse
profile might be distorted by a drift in the \nustar\ clock, causing a
smearing of the pulse, and therefore a decrease in the measured pulsed
fraction.  The attempt to correct for the \nustar\ clock drift is
discussed in Section~\ref{sec:clockdrift} and the smearing of the
pulse is quantified in Section~\ref{sec:smearing}.

\subsection{Searching and correcting for the residual clock drift}
\label{sec:clockdrift}

The \nustar\ onboard photon time stamping clock is known to suffer
drifts that are due to temperature variations as the telescope orbits
the Earth and is exposed to a variable environment.  The largest
fraction of this clock drift can be corrected for during the
processing of \nustar\ data with up-to-date \emph{clock
  files}\footnote{The \nustar\ clock file v050 was used in this
  analysis.} provided by the \nustar\ calibration team.  However,
there remains a residual clock drift.  Therefore, while the
\nustar\ temporal resolution is $2\unit{\mu s}$, inaccuracies in the
time stamping due to the clock drift limit the timing accuracy to
$\sim 0.5\msec$.

\begin{figure}
  \centering
  \makebox[0cm]{\includegraphics[width=8.5cm]{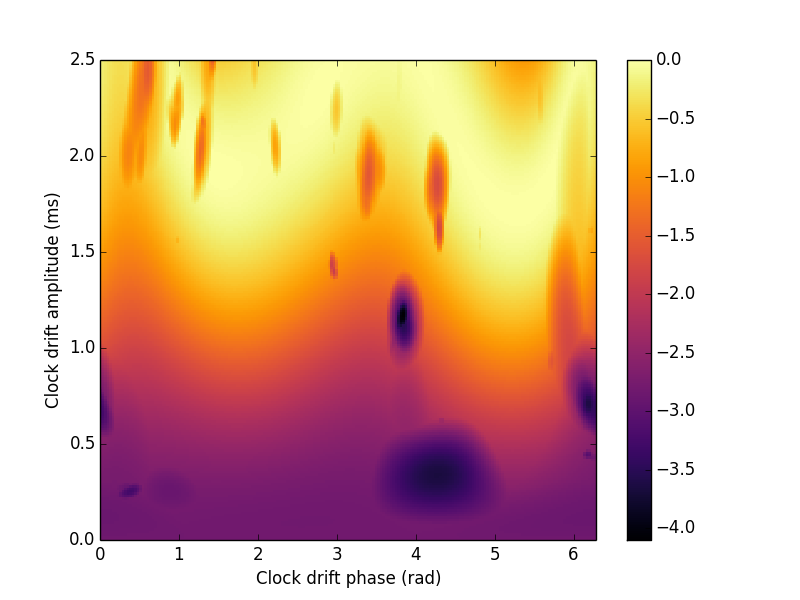}}
  \caption{Results of the search for the \nustar\ clock drift using
    the pulse profile of PSR~J0437--4715.  The clock drift is assumed
    to be sinusoidal, with unknown phase and amplitude, and with a
    period equal to that of the \nustar\ orbit.  The color scale
    corresponds to the H-score log-probability of such a score
    happening by chance (see Section~\ref{sec:clockdrift}).  }
  \label{fig:clockdrift}
\end{figure}

This residual clock drift has been noticed in observations of the Crab
pulsar \citep{madsen15}.  Residuals in the timing exhibited
modulations at a period of $\sim 97\unit{min}$, the satellite's
orbital period.  The detection of this residual clock drift in the
timing analysis of the Crab pulsar was only possible because this
source is bright enough that pulse time-of-arrivals can be obtained on
timescales of $\sim 5\unit{min}$, much shorter than the $\sim
97\unit{min}$ orbital period.  The residual drifts of the photon time
stamps cause a smearing of the otherwise sharper pulse.  While this
did not significantly affect the pulse shape of the Crab pulsar (spin
period $P=33\msec$), the effects on a faint 5.76\msec\ pulse could be
more dramatic.

At the count rates encountered in this \nustar\ observation of J0437
($\sim 800$ counts after background subtraction), the pulse shape
could be severely affected by the clock drift.  In an attempt to
recover the true, unaltered, pulse shape of J0437, a search in
amplitude and phase was performed for the residual clock drift,
assuming that it follows a sinusoidal pattern.  The exact amplitude
(in milliseconds) of the time correction that needs to be applied to
each photon is not well known; it was measured to be
$\sim0.4\pm0.1\msec$ from the Crab pulsar \nustar\ data
\citep{madsen15}.  In addition, the phase of the clock-drift
modulation is unknown.  Consequently, a grid search method in
amplitude-phase ($A$,$\phi$) was employed.  For each pair of
parameters, the following correction to the time stamp $t_{i}$ of each
photon $i$ was applied:
\begin{equation}
\label{eq:deltatime}
	\Delta t_{i} = A \sin\left( \frac{2\pi t_{i}}{P_{\rm orb}}+
        \phi \right) ,
\end{equation}
where $P_{\rm orb}$ is the satellite orbital period at the time of the
observation, obtained from the orbit files.  The time correction
$\Delta t_{i}$ can be converted into a correction $\Delta \theta_{i}$
to the phase $\theta_{i}$ of each photon, according to:
\begin{equation}
\label{eq:deltatheta}
	\Delta \theta_{i} = \frac{\Delta t_{i}}{P_{t_{0}, {\rm J0437}}
          + \dot{P}_{t_{0}, {\rm J0437}}\left( t_{i} - t_{0} \right)},
\end{equation}
where $P_{\rm J0437}$ and $\dot{P}_{\rm J0437}$ are the period and
period derivative of J0437 at the reference time $t_{0}$ (from the
ephemeris).

Each pair of parameters ($A$,$\phi$) therefore resulted in a corrected
profile in which the significance of a pulse was investigated using an
H-test.  From this search, a grid of H values was obtained, or
equivalently, H-test probabilities that allows one to determine
whether the corrections applied by each pair of parameters result in a
detected/improved pulse for J0437.  The clock drift amplitudes and
phases were sampled in the ranges 0--2.5\msec\ (step of 0.0125\msec)
and $0-2\pi$ (step of $\pi/200$), respectively.
Figure~\ref{fig:clockdrift} shows the H-test probabilities obtained
from the grid-search.  Low H-test probabilities (dark purple)
correspond to peaks in the H-score.  We found that none of the
detected peaks gives an H-test probability significantly higher than
that for the zero amplitude clock-drift correction.  In particular,
the peak at $A\sim1.2\msec$ is likely to be a statistical coincidence
since there is no evidence from the Crab pulsar timing that the clock
drift amplitude is that large.  However, the less-significant peak at
$A\sim0.3\msec$ is closer to the expected residual clock drift
amplitude (see above, and \citealt{madsen15}), but the low
significance of the improvement prevents us from making any firm
conclusions.  More importantly, the $A\sim0.3\msec$ solution does not
change significantly the pulsed fraction or the morphology of the
pulse profile.

\begin{figure}
  \centering
  \makebox[0cm]{\includegraphics[width=8.2cm]{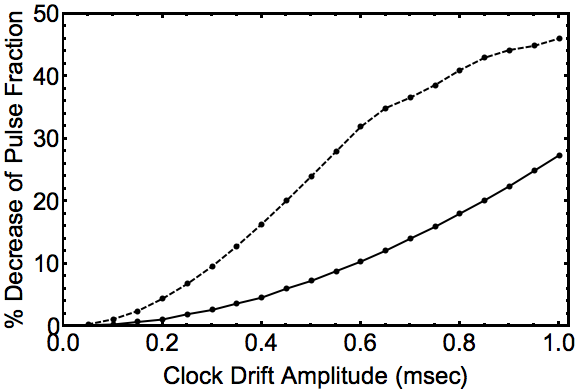}}
  \caption{Fractional difference between the intrinsic pulsed fraction
    of a 5.76\msec\ sinusoidal pulse profile and the pulsed fraction
    of a pulse profile distorted by the \nustar\ clock drift, as a
    function of the clock drift amplitude.  The solid line corresponds
    to a sinusoidal simulated pulse profile, and the dashed line
    corresponds to simulated narrow Gaussian pulse profile with a 10\%
    duty cycle ($1\sigma$ width).  For a clock drift amplitude of
    0.4\msec\ \citep{madsen15}, the measured pulsed fraction is
    artificially reduced by $\sim 5\%$, assuming a sinusoidal pulse
    profile, and by $\sim16\%$ for a short duty cycle profile.}
  \label{fig:PFvsAmp}
\end{figure}

\subsection{Effect of the residual clock drift on the pulsed fraction}
\label{sec:smearing}

To estimate the magnitude of the smearing effect on the observed pulse
profile, we simulated photons randomly drawn from a pulse profile
distribution (with $P=P_{\rm J0437}=5.76\msec$) over a total time
equal to the time span of our \nustar\ observation.  The randomly
drawn times of these photons were then modified by a small time
difference equal to what the clock drift would cause, following
Equation~\ref{eq:deltatime} (assuming $\phi=0.0$).  After folding
these clock-drift modified photon times at the period $P=5.76\msec$,
the pulsed fraction of this constructed pulse profile was
calculated. It was then compared to the pulsed fraction of the pulse
profile obtained before altering the photon times with the clock-drift
effect.  This process was repeated for a range of clock-drift
amplitudes between 0.0 and 1.0\msec, and 100 times for each amplitude.
Finally, we calculated the average difference between the intrinsic
pulsed fraction and that affected by a clock drift.  This simulation
was performed for a purely sinusoidal pulse profile, as well as for a
small duty cycle pulse profile -- we used a Gaussian profile with
$1\sigma=0.05 P_{\rm J0437}$.  The percent decreases of the pulse
fractions in these two cases are shown in Figure~\ref{fig:PFvsAmp}.
If the true residual clock drift of \nustar\ has an amplitude of
0.4\msec, as suggested by the observations of the Crab Pulsar
\citep{madsen15}, and by the clock-drift search described in
Section~\ref{sec:clockdrift}, the measured pulsed fraction is
underestimated by about $\sim 16\%$ in the case of a short duty cycle
pulse profile, and by $\sim 5\%$ in the case of a sinusoidal pulse
profile.  While this effect is currently buried in the statistical
uncertainties for J0437, it can become significant with deeper
observations, and for other bright MSPs with $P_{\rm spin}\approxlt
5\msec$. For example, faster pulsars, such as SAX~J1808.4--3658
($P_{\rm spin} = 2.5\msec$, \citealt{wijnands98}), observed with
\nustar, can expect to have their measured pulsed fraction biased by
$\sim40\%$ due to a 0.4\msec\ clock drift, assuming a small duty cycle
pulse profile.

\section{Spectral Analysis}
\label{sec:spectralAnalysis}

As was reported in previous work, the spectrum of J0437 contains
multiple components: two thermal components, and a non-thermal
component modeled with a PL
\citep{zavlin02b,zavlin06,durant12,bogdanov13}.  A third thermal
component (with $\kteff\sim40\eV$) has been suggested to fit the
lowest energy range available in the \xmm -MOS data
\citep{bogdanov13}.  Furthermore, \hstlong\ observations also seem to
advocate for a third thermal component representing the cold emission
from the NS surface, for which the Rayleigh-Jeans tail is detected in
the far UV and appears in excess of the modeling of the
white-dwarf companion atmosphere \citep{durant12}.  The
\nustar\ spectrum of J0437 complements the \xmm -MOS spectra up to
20\keV\ range, and provides a useful handle on the non-thermal
emission.  However, because the \nustar\ and \xmm\ observations were
obtained at a different epoch, the joint analysis described below
assumes that J0437 is a non-variable source.  This assumption is based
on the lack of evidence for long term variability of the thermal
emission in non-accreting MSPs like J0437 \citep{kaspi06}.

\subsection{\nustar\ spectroscopy}
\label{sec:nustar}
The analysis of the \nustar\ data was first performed without the \xmm
-MOS data.  In the 3--20\keV\ range covered by \nustar, the
non-thermal component dominates the emission, and a simple unabsorbed
PL spectral model was used.  Absorption of \xray s by the Galactic
interstellar medium in that energy range plays a negligible role.  A
multiplicative constant $c$, fixed to unity for the FPMA spectrum and
fitted for the FPMB spectrum, was added to take into account the
cross-correlation uncertainties between the two modules.  All
following parameters resulting from the spectral analyses are quoted
with 90\% confidence.  The photon index measured is
$\Gamma=1.60\pm0.25$, consistent with, but significantly restricting
the range of values previously tested or reported
\citep{durant12,bogdanov13}.  The multiplicative constant is
consistent with unity, $c=1.17\ud{0.28}{0.23}$.  Adding a hot
$kT=260\eV$ blackbody (as observed in the \xmm -MOS data analysis of
\citealt{bogdanov13}) with a fixed temperature and free normalization
did not result in any significant change in the photon index.

\begin{figure}
  \centering
  \makebox[0cm]{\includegraphics[width=8.5cm]{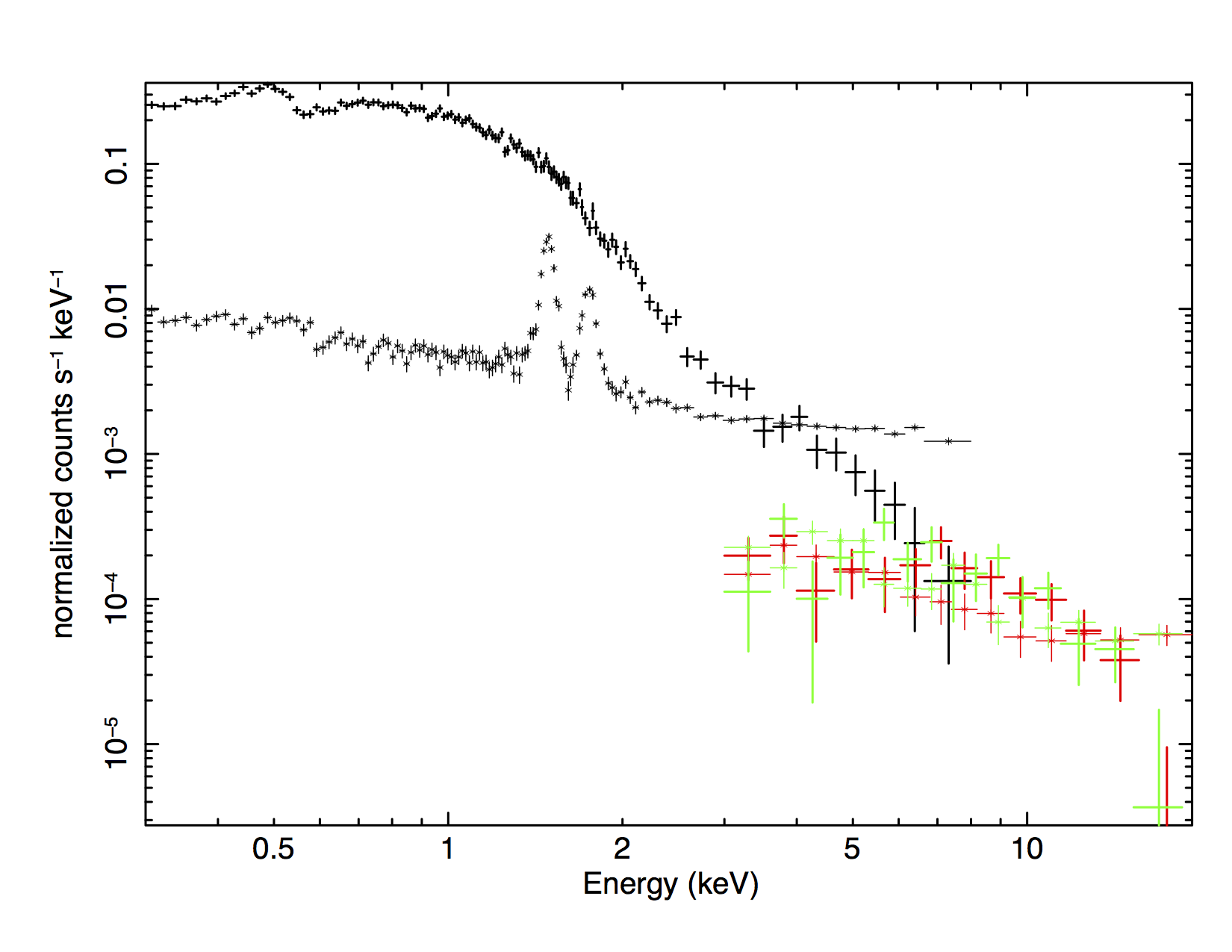}}
  \caption{\xmm-MOS (black), \nustar-FPMA (red) and \nustar-FPMB
    (green) count spectra of PSR~J0437--4715 (data points with thick lines),
    together with their respective background spectra ('$\times$' data
    points with thin error bars).}
  \label{fig:bkgspectra}
\end{figure}

\subsection{Joint spectral analysis}

This subsection concerns the joint fit of the \xmm -MOS and
\nustar\ spectra.  Figure~\ref{fig:bkgspectra} shows the count spectra
obtained with \xmm -MOS and \nustar, together with their respective
backgrounds.

In the previous work, the low S/N of the \xmm -MOS data at energies
above $\sim3\keV$ allowed the measured PL component to have indices in
the range $\sim0.8-2.9$ at the 90\% confidence level
\citep{durant12,bogdanov13}, depending on the number and types of the
thermal components fitting the soft \xray\ emission.  Specifically,
two or three thermal components both led to statistically acceptable
fits, but resulted in different PL photon indices.  Omitting the third
(coldest) thermal component could be compensated for by a soft PL
(higher $\Gamma$) which would dominate the very-soft emission and
account for the excess of counts in the softest energy band
($<0.3\keV$).  We note that this emission at $E<0.3\keV$ is detected
in the \xmm -MOS data \citep{bogdanov13}, but in an energy band where
the spectral calibration is undetermined.

\begin{figure}
  \centering
  \makebox[0cm]{\includegraphics[width=9.0cm]{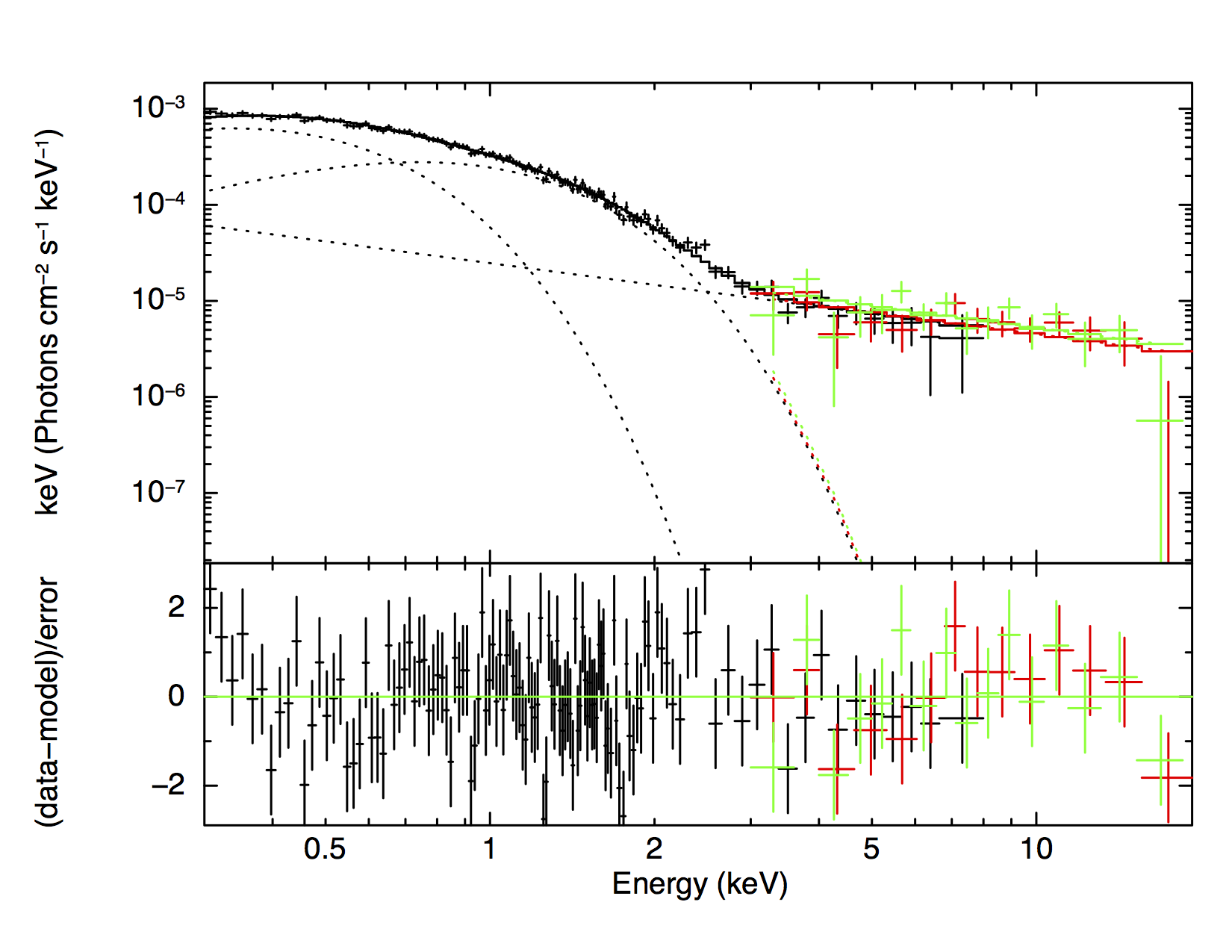}}
  \caption{\xmm-MOS (black), \nustar-FPMA (red) and \nustar-FPMB
    (green) unfolded spectra of PSR~J0437--4715 fitted with an
    absorbed double-blackbody plus power-law model.  While the fit is
    statistically acceptable, some structure in the residuals can be
    seen below 1\keV.}
  \label{fig:2BB+POW}
\end{figure}

Constraints on the non-thermal emission from the \nustar\ data
therefore permit differentiation between the range of possible soft
\xray\ emission models proposed.  The \nustar\ spectrum was
simultaneously fit with the \xmm -MOS spectrum.  As performed above, a
multiplicative constant $c$ was used to account for cross-correlation
uncertainties between the MOS (fixed $c$) and the FPMA/B spectra (one
fitted $c$ parameter for each module).  In this section, blackbody
components were used for the thermal emission.  Section~\ref{sec:ccl}
discusses the use of NS atmosphere models to describe the hot spot
emission and demonstrates that the choice of atmosphere model or
Planck spectrum does not significantly affect the measured PL index.
Finally, the Galactic absorption was quantified with the column
density of hydrogen \nh\ (noted \nht\ thereafter, when expressed in
units of $\ee{20}\unit{atoms\percmsq}$). It was modeled using the
\emph{phabs} model, with the \emph{bcmc} cross-sections
\citep{balucinska92} and the \emph{wilm} abundances \citep{wilms00}.
Alternative abundances have also been tested, as well as the
\emph{tbabs} absorption model.  The fits were insensitive to these
changes: all parameters were consistent well within the 1$\sigma$
uncertainties of the values obtained with \emph{phabs} and
\emph{wilm}.

\begin{table*}
 \centering
 %\begin{minipage}{140mm}
 \caption{Phase-averaged Spectral Analysis of PSR~J0437--4715}
 \resizebox{\linewidth}{!}{%
   \begin{tabular}{c|ccccccccccc}
     \hline
      Spectral & \nht & $kT_{1}$ & $R_{\rm eff, 1}$ & $kT_{2}$ & $R_{\rm eff, 2}$ & $kT_{3}$ & $R_{\rm eff, 3}$ & $\Gamma_{\rm PL}$ & N$_{\rm PL}$\,$^{a}$  & \chisqnu/dof (n.h.p) & $f$-test\\
      Model    &      & (eV)            &   (km)     &     (eV)        &    (km)    &        (eV)      &    (km)   &                  &  &                      & probability $^{b}$\\
     \hline
     \multicolumn{12}{c}{\bf \nustar\ FPMA/B} \\
     PL$^c$ & -- & -- & -- & -- & -- & -- & -- & 1.60\ppm 0.25 & 1.9\ud{1.3}{0.8} & 1.02/27 (0.43) & -- \\
     \hline
     \multicolumn{12}{c}{\bf \xmm-MOS1/2 + \nustar\ FPMA/B} \\
     2 BB+PL$^d$\ & $<$ 0.3 & --            & -- & 120\ppm 6 &  0.25\ud{0.03}{0.02} & 262\ppm 10 & 0.051\ud{0.006}{0.005} & 1.75\ppm 0.25 & 2.5\ud{1.2}{0.8} & 1.14/147 (0.13) & -- \\
     3 BB+PL$^e$\ & $<$ 8.9 & 34\ud{54}{14} & 11\ud{55}{7}   & 124\ud{50}{12} & 0.24\ud{0.21}{0.13}& 270\ud{40}{13} & 0.046\ud{0.009}{0.016} & 1.65\ppm 0.25 & 2.1\ud{1.0}{0.7} & 1.04/145 (0.34) & 0.0009 \\
     \hline
     \multicolumn{12}{c}{\bf \rosat-PSPC + \xmm-MOS1/2 + \nustar\ FPMA/B} \\
     2 BB+PL$^d$\ & $<$ 0.6 & -- & -- & 111\ud{5}{6} & 0.26\ud{0.05}{0.09} & 250\ppm 7 & 0.056\ud{0.006}{0.005} & 2.15\ud{0.55}{0.20} & 4.5\ud{4.6}{0.4} & 1.38/175 (0.0006) & -- \\
     3 BB+PL$^e$\ & 2.4\ud{1.3}{1.2} & 32\ud{7}{5} & 11\ud{23}{5} & 125\ud{10}{9} & 0.25\ud{0.06}{0.04} & 271\ud{14}{12} &  0.046\ud{0.007}{0.006} & 1.65\ppm 0.24 & 2.1\ud{0.9}{0.7} & 1.02/173 (0.41) &  2\tee{-12}\\
     2 {\rm Hatm}+BB+PL$^f$ & 3.3\ud{1.6}{1.3} & 27\ud{5}{4} & 10--100 & 51\ppm6 & 4.5\ud{2.3}{1.7} & 147\ud{11}{8} & 0.4\ppm0.1 & 1.50\ppm0.25 & 1.4\ud{0.7}{0.5} & 1.00/173 (0.48) & -- \\
     \hline
     \hline
     \multicolumn{12}{l}{$^a$\, Power-law normalization in units of $\ee{-5}\unit{photons\,keV^{-1}\percmsq\persec}$.}\\
     \multicolumn{12}{l}{$^b$\, $f$-test probability of adding an additional thermal component, obtained by comparing the \chisq\ fit to the fit directly above.}\\
     \multicolumn{12}{l}{$^c$\, Spectral model used: {\tt phabs$\times$powerlaw}.}\\
     \multicolumn{12}{l}{$^d$\, Spectral model used: {\tt phabs$\times$(blackbody+blackbody+powerlaw)}.}\\
     \multicolumn{12}{l}{$^e$\, Spectral model used: {\tt phabs$\times$(blackbody+blackbody+blackbody+powerlaw)}.}\\
     \multicolumn{12}{l}{$^f$\, Spectral model used: {\tt phabs$\times$(blackbody+nsatmos+nsatmos+powerlaw)}.  The {\tt nsatmos} components assumed \rns=13.5\km\ and \mns=1.44\msun\ (see text). }\\
     \multicolumn{12}{l}{The $R_{\rm eff}$ values are deduced from the fitted normalization parameters of the {\tt nsatmos} model.}\\
     \label{tab:PhaseAve}
   \end{tabular}
 }
 %\end{minipage}
\end{table*}

As found in the previous work, the models {\tt blackbody+powerlaw} and
{\tt blackbody+blackbody} are not statistically acceptable fits to
present data.  The former, with a null hypothesis probability
$p\sim\ee{-6}$, presents highly structured residuals, especially above
$6\keV$.  The latter has $p\sim\ee{-45}$, and does not fit any of the
high energy emission above $3\keV$.  Considering the spectral model
{\tt blackbody+blackbody+powerlaw} (2BB+PL hereafter) resulted in a
statistically acceptable fit (Figure~\ref{fig:2BB+POW}), with
\Chisq{1.13}{147}{0.13}, and with a PL photon index
$\Gamma=1.75\pm0.25$.  In this case, the hydrogen column density
remains unconstrained, $\nht<0.3$, and consistent with zero.

The two multiplicative constants used for the two \nustar\ spectra are
consistent with unity: $c_{\rm FPMA}=1.03\ud{0.25}{0.21}$ and $c_{\rm
  FPMB}=1.20\ud{0.28}{0.24}$.  Because of the low S/N, the
multiplicative constants have large error bars, but they are
nonetheless consistent with those obtained with the high-S/N data of
\nustar\ calibration sources \citep{madsen15}.  To investigate the
possibility that the large value of $c_{\rm FPMB}$ is skewing the
results, we performed a fit by fixing the multiplicative constant
between FPMA and FPMB to 1.03, the typical value measured with the
calibration sources \citep{madsen15}; a single free multiplicative
constant between \xmm\ and the \nustar\ detectors is still fitted to
account for cross-calibration uncertainties between the two
observatories.  In this exercise\footnote{We performed the same
  exercise on the \nustar\ data alone (Section~\ref{sec:nustar}).  The
  PL index was the same, and the PL normalization was 5\% larger, but
  still consistent with that of Table~\ref{tab:PhaseAve}.}, the
parameters changed by $<0.5\%$, i.e., they are all consistent well
within the uncertainties presented in Table~\ref{tab:PhaseAve}.  This
indicates that the results are not skewed by the fitted multiplicative
constants.

The temperatures of the two blackbodies obtained with the 2BB+PL model
are consistent with those of the two hottest blackbodies in the three
blackbody case of \cite{bogdanov13}.  It is important to note that in
the present analysis, the 0.1--0.3\keV\ range, where the coldest
thermal component dominates, is excluded due to the lack of
calibration of the \xmm-MOS detector below $0.3\keV$.

\begin{figure}
  \centering
  \makebox[0cm]{\includegraphics[width=9.0cm]{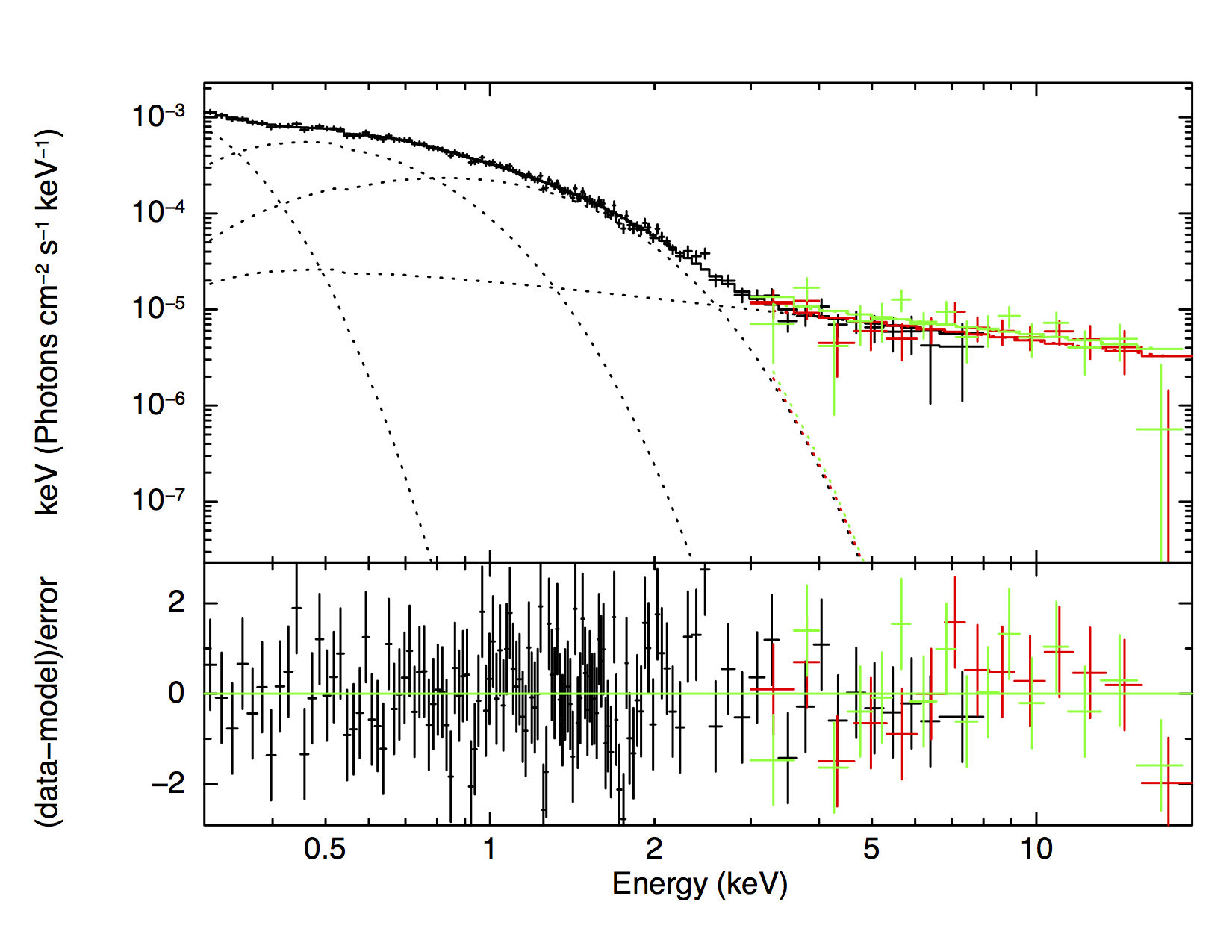}}
  \caption{\xmm-MOS (black), \nustar-FPMA (red) and \nustar-FPMB
    (green) unfolded spectra of PSR~J0437--4715 fitted with an
    absorbed triple-blackbody plus power-law model.  The addition of a
    third thermal component improves the fit ($f$-test probability of
    0.0009), and improves the residuals below 1\keV\ compared to
    Figure~\ref{fig:2BB+POW}.}
  \label{fig:3BB+POW}
\end{figure}

Nonetheless, adding a third thermal component in the present analysis
improved the statistics: \Chisq{1.04}{145}{0.34}.  The $f$-test gives
a probability of 0.0009 of the \chisq\ improvement happening by chance
when adding the thermal blackbody.  The temperature of this added
component is $\kteff=34\ud{54}{14}\eV$, consistent with the value
reported previously \citep{bogdanov13} using the MOS data down to
$0.1\keV$.  This new fit improves the residuals in the lowest energy
bins: 0.3--0.35\keV\,(Figure~\ref{fig:3BB+POW}), but results in a
poorly constrained temperature because of the limited energy range
where this cold thermal component dominates.  Therefore, this fit
indicates that the third thermal component, emitted by the NS surface,
likely exists.  However, because of the lack of calibration, it is not
recommended to include the 0.1--0.3\keV\ MOS data to better constrain
this cold thermal emission and draw any firm conclusions.

\begin{figure}
  \centering
  \makebox[0cm]{\includegraphics[width=9.0cm]{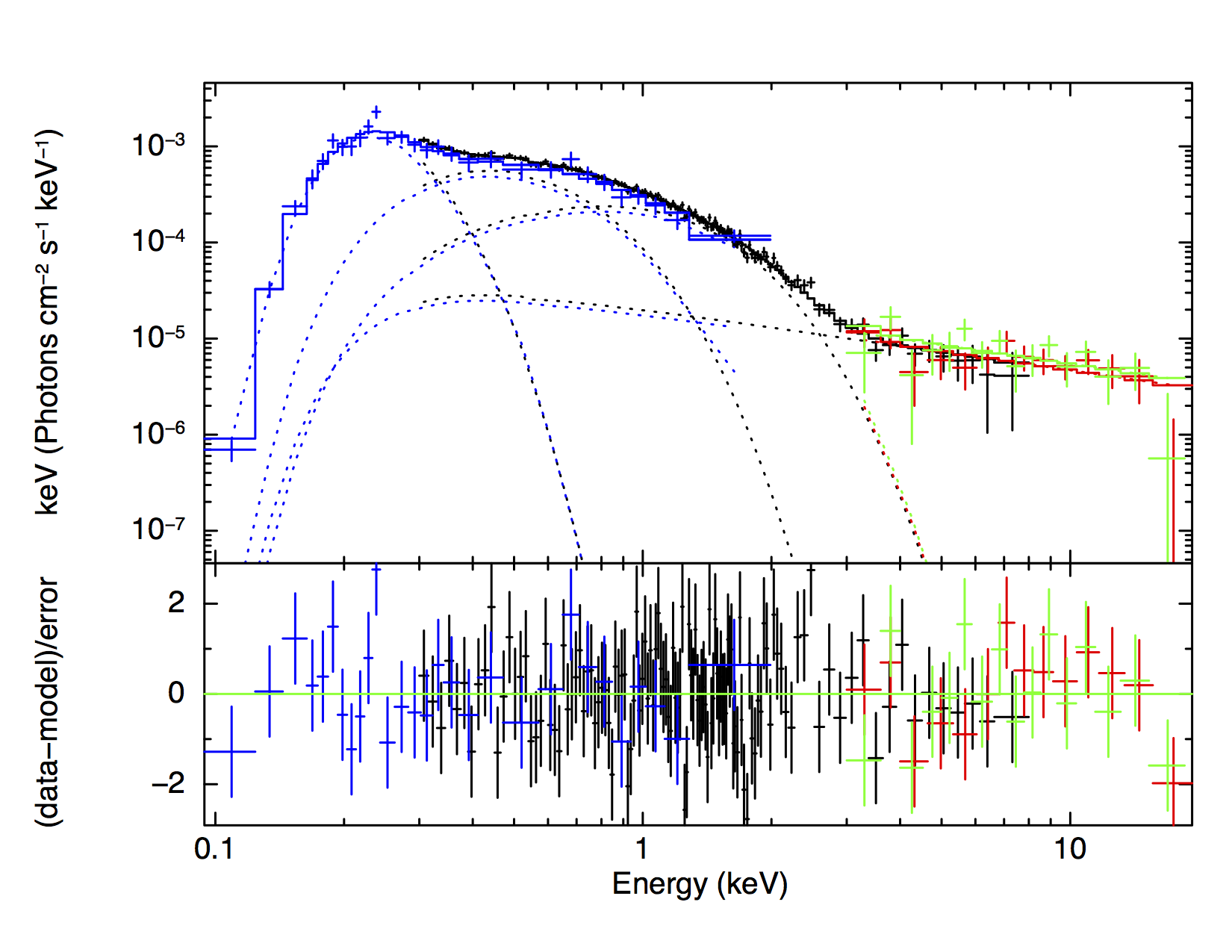}}
  \caption{\xmm-MOS (black), \nustar-FPMA (red), \nustar-FPMB (green)
    and \rosat -PSPC (blue) unfolded spectra of PSR~J0437--4715 fitted
    with an absorbed triple-blackbody plus power-law model.  The
    addition of the \rosat\ data permits us to obtain better
    constraints on the coldest of the three thermal components and on
    the amount of absorption.}
  \label{fig:ROSAT3BB+POW}
\end{figure}

To further investigate this, the \rosat-PSPC data, and its calibrated
spectral information down to 0.1\keV, proved a useful addition to the
joint spectral analysis presented here.  An additional multiplicative
constant was added to account for the cross-calibration uncertainties
of the \rosat-PSPC detectors with \xmm-MOS and \nustar-FPMA/B.  The
combined \rosat-PSPC, \xmm-MOS and \nustar\ spectra could not be
adequately fit by a 2BB+PL model ($p\sim 0.0006$, see
Table~\ref{tab:PhaseAve}).  The excess emission in the
0.1--0.3\keV\ band added by the \rosat\ data was not adequately fitted
by the two thermal components and was therefore compensated by a soft
PL ($\Gamma=2.15\ud{0.55}{0.20}$).  This in turn left a large excess
of counts in the \nustar\ band, causing the observed increase in the
\chisq\ statistic.

However, adding a third thermal component confirmed what was observed
with the \xmm-MOS and \nustar\ data.  Specifically, the
0.1--0.3\keV\ emission is well fitted by a blackbody with
$\kteff=32\ud{7}{5}\eV$.  Overall, this leads to an acceptable fit
statistic \Chisq{1.02}{173}{0.41}, and without clear structure in
residuals (see Figure~\ref{fig:ROSAT3BB+POW}).  Furthermore, while the
amount of absorption was unconstrained (upper limit on $\nh$) with the
\xmm-MOS and \nustar\ data, the inclusion of the \rosat-PSPC spectrum
permitted the measurement of the Galactic absorption, $\nht =
2.4\ud{1.3}{1.2}$, without ambiguity on the rest of the emission,
especially at high energy.

In this spectral analysis, the \xmm-pn data were not included, since
despite their high S/N, they do not significantly improve the
constraints on the thermal components.  For example, adding the
\xmm-pn data to the \xmm-MOS and \nustar\ spectra resulted in
blackbody temperatures $\kteff=117\pm 6\eV$ and
$\kteff=254\ud{9}{8}\eV$ when fitting a 2BB+PL model.  These values
are consistent with the \xmm-MOS/\nustar\ spectral fits (see
Table~\ref{tab:PhaseAve}), but more importantly, the improvements on
the statistical uncertainties of the measurements are negligible.
Indeed, the timing mode of the \xmm-pn data has a limited reliable
energy range, and requires an additional cross-correlation coefficient
(multiplicative constant) which limits the impact of the additional
S/N.

\vspace{-0.5cm}

\subsection{Phase-resolved spectral analysis}
\label{sec:PhaseRes}

Using the phase-folded lightcurve presented in
Section~\ref{sec:timing}, the on-pulse and off-pulse spectra were
extracted (see Figure~\ref{fig:pulse}).  Because of the limited number
of counts in the \nustar\ observations of J0437, the analysis was
restricted to just two phase bins.  Just as performed for the
\nustar\ data alone, a simple PL model was used for the on- and
off-pulse spectra, together with a multiplicative constant accounting
for cross-calibration uncertainties.  For the on-pulse spectra, a PL
photon index $\Gamma_{\rm on}=1.6\pm0.3$ was obtained, while the
off-pulse photon index was $\Gamma_{\rm off}=1.5\pm0.3$.  Both fits
are statistically acceptable (null hypothesis probability of
$p\sim0.2$ and $p\sim0.8$, respectively).

The phase-resolved spectral analysis does not reveal a significant
change in the PL index, given the count statistics available.
Higher S/N data will be necessary to draw any firm
conclusions regarding the variation with phase of the high-energy
spectral shape of J0437.

\vspace{-0.5cm}

\section{Discussion and Conclusions}
\label{sec:ccl}

Using the well measured radio ephemeris of PSR~J0437--4715, pulsations
were observed with a significance of 3.7$\sigma$ in a
200\ksec\ \nustar\ observation, despite a spin period small enough to
potentially be affected by the \nustar\ clock drift modulated at the
satellite orbital period.  Under the assumption that the residual
clock drift is sinusoidal over an orbit of the satellite, causing a
smearing of the pulse profile, we attempted to recover and correct for
a sinusoidal clock drift of unknown phase and amplitude.  However, no
particular pair of phase and amplitude was found to significantly
improve the pulse profile.  The measured high energy pulsed fraction
of J0437 is $24\pm6\%$ (2--20\keV).  When separated into two energy
bands, the pulsations are marginally detected, likely due to the low
S/N in each band. The pulse fractions ($32\pm9\%$ in 2--6\keV, and
$20\pm8\%$ in 6--20\keV) show hints of a decrease with energy,
although not significantly.  However, it is important to keep in mind
that the pulsed fractions of MSPs, like J0437, observed with
\nustar\ are probably underestimated due to the distortion of the
pulse profile by the clock drift.  The artificial decrease of the
pulse fraction depends on the pulse period and on the magnitude of the
clock drift (Figure~\ref{fig:PFvsAmp}).

The \nustar\ observation presented here provided the high-energy
constraint needed to remove ambiguities in the soft \xray\ spectral
modeling.  The S/N of previously available spectra was too low above
$\sim3\keV$ to fully charactetize the high energy emission.  While its
presence was confirmed and modeled with a PL in an
\xmmlong\ observation \citep{bogdanov13}, the photon index was poorly
constrained which resulted in uncertainties in the modeling of the
remainder of the soft \xray\ emission.  We showed that the
\nustar\ observations alone constrained the photon index of the
3--20\keV\ emission to $\Gamma= 1.60\pm0.25$, firmly excluding PL
photon indices of $\sim 2.5$, as sometimes obtained in fits of the
\xmmlong\ spectra alone (depending on the other spectral components
chosen).

With this new high-energy observation, and by comparing the 2BB+PL and
3BB+PL models, we demonstrated that the presence of a third thermal
component ($kT_{\rm BB}=32\ud{7}{5}\eV$) is required by the
\xray\ emission of J0437.  This cold component is thought to be
emanate from the entire surface of the NS, while the two hotter
thermal components describe the emission from the polar caps of the
rotating NS.  Such a result is in agreement with the far UV
emission which required a NS surface thermal component, since the
UV emission could not be accommodated by the white-dwarf
atmosphere emission \citep{durant12,kargaltsev04}.  With the 3BB+PL
spectral model, we found a PL index of $\Gamma= 1.65\pm0.24$.

\begin{figure}
  \centering
  \makebox[0cm]{\includegraphics[width=9.0cm]{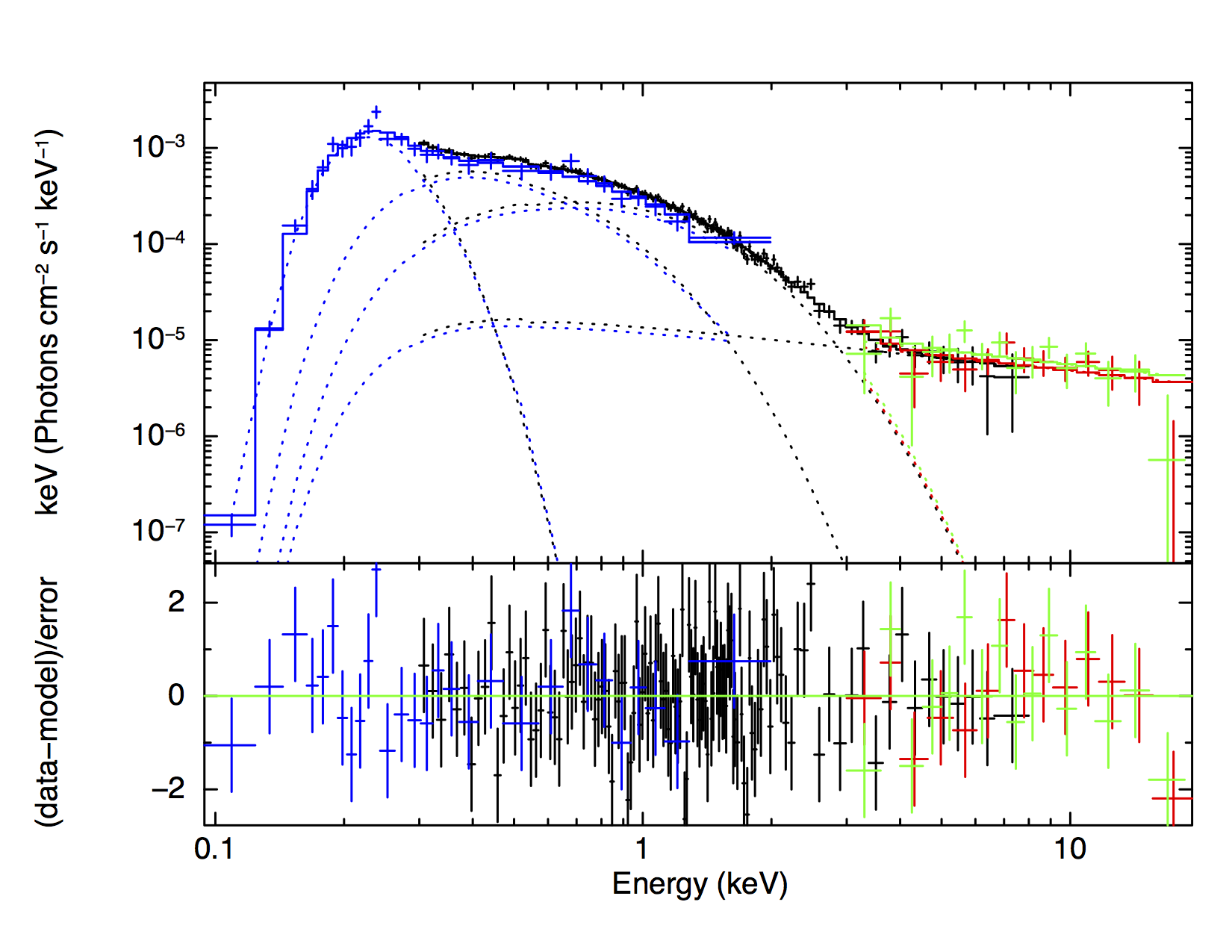}}
  \caption{\xmm-MOS (black), \nustar-FPMA (red), \nustar-FPMB (green)
    and \rosat -PSPC (blue) unfolded spectra of PSR~J0437--4715 fitted
    with one blackbody, two NS atmosphere {\tt nsatmos} components and
    a power law, together with a model for Galactic absorption.
    Replacing the Planck spectral components of
    Figure~\ref{fig:ROSAT3BB+POW} by {\tt nsatmos} component does not
    significantly change the measured PL index, but it increases the
    contribution of the thermal emission to the flux above $3\keV$.}
  \label{fig:ROSAT2NSATMOS+POW}
\end{figure}

NS atmosphere models and their beamed emission patterns offer a more
realistic representation of the surface thermal emission of NSs than
an isotropically radiating Planck spectrum
\citep[e.g.,][]{bogdanov07}.  The three Planckian components were
therefore replaced by three realistic NS hydrogen atmosphere models
({\tt nsatmos}, \citealt{heinke06a}).  We found an acceptable fit and
a PL index $\Gamma= 1.46\pm0.25$, consistent with the 3BB+PL
case. However, the coldest \emph{nsatmos} has a temperature pegged at
the lower limit allowed by the model ($\log(T/\unit{K})=5.0$),
preventing us to properly estimate the errors on the surface
temperature.  The coldest thermal component was replaced by a {\tt
  blackbody} component, and we obtained an acceptable fit with a PL
index $\Gamma= 1.50\pm0.25$ (see Table~\ref{tab:PhaseAve}, for the
full set of parameters).  Because the high energy tail of NS
atmosphere models has slightly more flux than a Planck spectrum
\citep[see for example,][]{zavlin96,heinke06a,haakonsen12}, the PL
index measured is slightly harder than that measured in the 3BB+PL
case.  However, this change is not significant given the S/N available
here.

At $6\keV$, the hottest blackbody (in the case the 3BB+PL case) is a
few orders of magnitude fainter than the PL component (see
Figure~\ref{fig:ROSAT3BB+POW}).  However, when {\tt nsatmos} models
are used, the hottest component contributes to $\sim 0.3\%$ of the
flux at $6\keV$ (Figure~\ref{fig:ROSAT2NSATMOS+POW}) since it falls
off less steeply than a blackbody component.  Therefore, it cannot be
excluded that the pulsed emission below $6\keV$ is due, in part, to
the thermal emission.  However, above $6\keV$, the pulsed emission is
unlikely to originate from the thermal components since its
contribution is essentially negligible at 6\keV\ and above.  It is
important to keep in mind that the detection significance of the
pulsations is marginal when we split the observed photons into low
(2--6\keV) and high (6-20\keV) energy bands (see
Section~\ref{sec:timing}).

Finally, in the phase-resolved spectroscopic analysis of the
\nustar\ data, there is no evidence of any significant change in the
PL index, most likely due to the low S/N of the spectra obtained when
splitting into two phase bins.

When comparing to other non-accreting MSPs\footnote{We used the list
  of MSPs detected by \fermi\ \citep[Table 2 of][]{marelli11}, as well
  as other MSPs not detected by \fermi, such as PSR~B1937--21, and
  MSPs in the globular clusters 47~Tuc \citep{bogdanov06a} and
  NGC~6397 \citep{bogdanov10}.}, the measured PL of J0437 is
consistent with that measured for most of them.  The MSP
PSR~J1024--0719 has a photon index $\Gamma=3.7\pm0.8$ when fitted with
a single PL model \citep{zavlin06}.  However, a model with a hot
thermal polar cap plus a PL does not provide a constraint on the PL
photon index for that MSP.  For the MSP PSR~B1937+21, the PL photon
index is $\Gamma=0.9\pm0.1$, and the fit is not significantly improved
when adding a thermal component \citep{ng14}, providing only an upper
limit on the temperature.  The MSP PSR~J0030$+$0451 is quite analogous
to J0437, and when fitted with two NS atmosphere models plus a PL, the
measured photon index is $\Gamma=2.0\pm0.2$, consistent with that of
J0437, given the uncertainties on both measurements.  However, if a
single thermal component plus PL model, or two blackbodies plus PL
model are chosen, the photon index $\Gamma=3.1\pm0.1$ is not
consistent with that of J0437.  This indicates that comparing the PL
photon indices of these non-accreting MSPs is rather delicate.  In the
soft X-ray pass band, the available S/N of these sources is generally
rather weak, J0437 (the nearest MSP) being the brightest, which leads
to poorly constrained spectral parameters.  More importantly, the
photon index of the PL components depends strongly on the overall
spectral model, and it is frequently found for many X-ray faint MSPs
that a blackbody, a PL, or the sum of both, fit the soft \xray\ data
equally well.

The PL photon index that we found for J0437 is consistent with the
photon flux detected by \fermi\ above 0.1\unit{GeV} when extending to
gamma-ray energies:
$(4.4\pm0.1)\tee{-8}\unit{photons\percmsq\persec}$. However, this may
simply be a coincidence since the gamma-ray emission and the hard
\xray\ non-thermal emission have different origins.  Curvature
radiation is responsible for the gamma-ray emission
\citep{cheng86,romani96,harding08,lyutikov13}, while the non-thermal
\xray\ emission may be due to synchrotron radiation from the shock
caused by the interaction of the pulsar wind and its companion wind
\citep{arons93}.  However, this possibility would be excluded if the
likely pulsations of the non-thermal emission above $>6\keV$ are
confirmed, since synchrotron radiation would not produce
pulsed emission.  Furthermore, it had been shown that the non-thermal
\xray\ emission of J0437 is lower than what would be expected from a
shock given the binary system orbital properties
\citep{bogdanov06b}. The proposed alternative interpretation of the
non-thermal emission involves electrons and positrons in the pulsar
magnetosphere causing the weak Comptonization of the thermal photons
\citep{bogdanov06b}.  This interpretation could explain that the
non-thermal emission of J0437 is likely pulsed as observed in the
present work.  However, the existence of such pulsations would depend
on the viewing geometry, and on the location from which scattered
photons originate.  Such information could be extracted from a higher
S/N pulse profile at energies above 6\keV\ where the thermal emission
is negligible.

The non-thermal emission characterized in this work confirms what had
been observed in a previous work where a PL with $\Gamma=1.56$ matches
both the \xmm\ emission $\approxlt8\keV$ and at \fermi\ energies
\citep[Figure 13 of][]{durant12}. Furthermore, when extended to UV
energies, the PL emission complements the ``cold'' surface emission to
fit the UV emission detected by the \hstlong\ \citep{durant12}.  The
present \nustar\ observation and the measurements of the PL photon
index shed light on the broadband emission of J0437, confirming that
the non-thermal emission is likely to extend over many decades of
energy.  Furthermore, knowledge of the non-thermal emission allowed us
to constrain the temperature of the NS cold surface emission as well
as the amount of \xray\ absorption (described by \nh).  Therefore, it
favours the cut-PL model (with a cut-off at $\sim 1.1\unit{GeV}$),
rather than a variable PL model \citep{durant12}.

Overall, a global understanding of the broadband emission of
J0437, and in particular a clear characterization of its thermal
emission, are crucial to extract the mass and radius of the pulsar.
J0437 is the key target for observations with \nicer, a NASA
experiment to be mounted on the International Space Station circa
2017.  The modeling of the pulse profiles and of the spectral emission
obtained from combining 1\unit{Msec} of observations with \nicer\ is
expected to produce stringent constraints on the compactness of J0437
($\mns/\rns$).  Since the mass of the pulsar is known from radio
timing observations, combining the \xray\ and radio measurements of
\mns\ and \rns\ would result in tight constraints in the mass-radius
parameter space, and therefore in stringent constraints on the dense
matter equation of state \citep{gendreau12}.  However, such an
analysis requires a full understanding of the emission in the
\nicer\ passband, i.e. 0.2--10\keV.

It is difficult to estimate quantitatively how this
\nustar\ observation will improve the $\mns/\rns$ measurements that
can be derived from currently available \xmm\ data, and from future
\nicer\ data.  Specifically, it is unclear how the characterization of
the non-thermal emission that was performed in the present work will
affect the modeling of the pulse profile.  Nevertheless, in the
previous work, the contribution of the non-thermal emission was
included as an un-pulsed contribution with a $\Gamma=1.56$ PL spectrum
\citep{bogdanov13}.  While we confirm here that the PL index is
consistent with $\Gamma=1.56$, we also demonstrated that the
non-thermal emission is probably pulsed, although this requires higher
S/N observations to be confirmed with higher certainty. This is
particularly important above 6\keV\ where the hot thermal component is
more than two orders of magnitude fainter and therefore does not
contribute to the pulsations.  Consequently, \nustar\ provides
information regarding the likely pulsed behaviour of the non-thermal
emission of J0437.  While no change in the PL index was detected
between the on- and off- pulse parts of the profiles, the varying flux
of the non-thermal emission as a function of phase ought to be
included in the modeling of the full pulse profile.

Even up to 3\keV\ (the upper limit of the range available for the
currently available \xmm-pn data), the PL contributes to the
phase-averaged flux (Figure~\ref{fig:ROSAT2NSATMOS+POW}), and
therefore contribute to the overall pulsed emission observed in the
\xmm\ data.  Note that in our best-fit model with {\tt nsatmos}
thermal components, the PL contribution to the flux becomes larger
than that of the hot thermal component at around $\sim2.7\keV$
(Figure~\ref{fig:ROSAT2NSATMOS+POW}).  By neglecting the pulsed
emission of the PL, one will attribute more pulsed flux to the thermal
components than in reality.  As a direct consequence of this, the
best-fit geometry of the model or the measured NS compactness could be
biased.

Confirming the existence of the third (coldest) thermal component and
measuring the temperature of the surface emission of J0437 is also of
great importance to confirm the existence of reheating mechanisms in
old NSs. Standard cooling models predict surface
temperatures of $<10^{4}\K$ after $10^{9}\yr$. However, evidence of a
$\sim 10^{5}\K$ surface temperature for J0437 despite its
characteristic age $>10^{8}\yr$ indicates that reheating mechanisms
might be at play inside old NSs \citep[for example,
  rotochemical heating, or frictional heating, see ][for more
  details]{gonzalez10}.  However, the present work used a blackbody to
fit the coldest thermal component, while a realistic atmosphere model
with a range of temperature compatible with $T<10^{5}\K$ should be
used. In addition, measurements of the UV fluxes of J0437
\citep[e.g.:][]{durant12} can be used to help constrain that coldest
thermal component since the Rayleigh-Jeans tail is detected in the UV
and Far UV range (González-Caniulef, Reisenegger \& Guillot, in
prep.).

In conclusion, future pulse profile and phase resolved spectral
modeling of J0437, in particular those of future \nicer\ data, must
take into account the likely pulsed non-thermal emission with a photon
index of $\Gamma=1.50\pm0.25$.  Ideally, the emission above
$\sim3\keV$ should be divided into more phase bins for a more accurate
determination of the non-thermal emission of J0437 as a function of
phase.  However, this would require higher S/N data, and therefore
deeper exposures with \nustar.

\smallskip

{\bf Acknowledgements} The authors thank the referee, Slavko Bogdanov,
for very useful suggestions that significantly improved this paper.
This work made use of data from the \nustar\ mission, a project led by
the California Institute of Technology, managed by the Jet Propulsion
Laboratory, and funded by the National Aeronautics and Space
Administration. The data analysis was performed with the \nustar\ Data
Analysis Software (NuSTARDAS) jointly developed by the ASI Science
Data Center (ASDC, Italy) and the California Institute of Technology
(USA).  S.G. is a FONDECYT post-doctoral fellow, funded by grant \#
3150428. VMK receives support from an NSERC Discovery Grant and
Accelerator Supplement, from the Centre de Recherche en Astrophysique
du Qu\'{e}bec, an R. Howard Webster Foundation Fellowship from the
Canadian Institute for Advanced Study, the Canada Research Chairs
Program and the Lorne Trottier Chair in Astrophysics and
Cosmology. Parts of this research were conducted by the Australian
Research Council Centre of Excellence for All-sky Astrophysics
(CAASTRO), through project number CE110001020.

\bibliographystyle{mnras} 
\bibliography{/Users/Sebastien/Research/CONTRIBUTIONS/Publications/biblio}

\begin{thebibliography}{}
\makeatletter
\relax
\def\mn@urlcharsother{\let\do\@makeother \do\$\do\&\do\#\do\^\do\_\do\%\do\~}
\def\mn@doi{\begingroup\mn@urlcharsother \@ifnextchar [ {\mn@doi@}
  {\mn@doi@[]}}
\def\mn@doi@[#1]#2{\def\@tempa{#1}\ifx\@tempa\@empty \href
  {http://dx.doi.org/#2} {doi:#2}\else \href {http://dx.doi.org/#2} {#1}\fi
  \endgroup}
\def\mn@eprint#1#2{\mn@eprint@#1:#2::\@nil}
\def\mn@eprint@arXiv#1{\href {http://arxiv.org/abs/#1} {{\tt arXiv:#1}}}
\def\mn@eprint@dblp#1{\href {http://dblp.uni-trier.de/rec/bibtex/#1.xml}
  {dblp:#1}}
\def\mn@eprint@#1:#2:#3:#4\@nil{\def\@tempa {#1}\def\@tempb {#2}\def\@tempc
  {#3}\ifx \@tempc \@empty \let \@tempc \@tempb \let \@tempb \@tempa \fi \ifx
  \@tempb \@empty \def\@tempb {arXiv}\fi \@ifundefined
  {mn@eprint@\@tempb}{\@tempb:\@tempc}{\expandafter \expandafter \csname
  mn@eprint@\@tempb\endcsname \expandafter{\@tempc}}}

\bibitem[\protect\citeauthoryear{{Abdo} et~al.,}{{Abdo} et~al.}{2009}]{abdo09}
{Abdo} A.~A.,  et~al., 2009, \mn@doi [Science] {10.1126/science.1176113}, \href
  {http://adsabs.harvard.edu/abs/2009Sci...325..848A} {325, 848}

\bibitem[\protect\citeauthoryear{{An} et~al.,}{{An} et~al.}{2015}]{an15}
{An} H.,  et~al., 2015, \mn@doi [\apj] {10.1088/0004-637X/807/1/93}, \href
  {http://adsabs.harvard.edu/abs/2015ApJ...807...93A} {807, 93}

\bibitem[\protect\citeauthoryear{{Arons} \& {Tavani}}{{Arons} \&
  {Tavani}}{1993}]{arons93}
{Arons} J.,  {Tavani} M.,  1993, \mn@doi [\apj] {10.1086/172198}, \href
  {http://adsabs.harvard.edu/abs/1993ApJ...403..249A} {403, 249}

\bibitem[\protect\citeauthoryear{{Balucinska-Church} \&
  {McCammon}}{{Balucinska-Church} \& {McCammon}}{1992}]{balucinska92}
{Balucinska-Church} M.,  {McCammon} D.,  1992, \mn@doi [\apj] {10.1086/172032},
  \href {http://adsabs.harvard.edu/abs/1992ApJ...400..699B} {400, 699}

\bibitem[\protect\citeauthoryear{{Becker} \& {Tr{\"u}mper}}{{Becker} \&
  {Tr{\"u}mper}}{1993}]{becker93}
{Becker} W.,  {Tr{\"u}mper} J.,  1993, \mn@doi [\nat] {10.1038/365528a0}, \href
  {http://adsabs.harvard.edu/abs/1993Natur.365..528B} {365, 528}

\bibitem[\protect\citeauthoryear{{Bogdanov}}{{Bogdanov}}{2013}]{bogdanov13}
{Bogdanov} S.,  2013, \mn@doi [\apj] {10.1088/0004-637X/762/2/96}, \href
  {http://adsabs.harvard.edu/abs/2013ApJ...762...96B} {762, 96}

\bibitem[\protect\citeauthoryear{{Bogdanov}, {Grindlay}, {Heinke}, {Camilo},
  {Freire}  \& {Becker}}{{Bogdanov} et~al.}{2006a}]{bogdanov06a}
{Bogdanov} S.,  {Grindlay} J.~E.,  {Heinke} C.~O.,  {Camilo} F.,  {Freire}
  P.~C.~C.,   {Becker} W.,  2006a, \mn@doi [\apj] {10.1086/505133}, \href
  {http://adsabs.harvard.edu/abs/2006ApJ...646.1104B} {646, 1104}

\bibitem[\protect\citeauthoryear{{Bogdanov}, {Grindlay}  \&
  {Rybicki}}{{Bogdanov} et~al.}{2006b}]{bogdanov06b}
{Bogdanov} S.,  {Grindlay} J.~E.,   {Rybicki} G.~B.,  2006b, \mn@doi [\apjl]
  {10.1086/507827}, \href {http://adsabs.harvard.edu/abs/2006ApJ...648L..55B}
  {648, L55}

\bibitem[\protect\citeauthoryear{{Bogdanov}, {Rybicki}  \&
  {Grindlay}}{{Bogdanov} et~al.}{2007}]{bogdanov07}
{Bogdanov} S.,  {Rybicki} G.~B.,   {Grindlay} J.~E.,  2007, \mn@doi [\apj]
  {10.1086/520793}, \href {http://adsabs.harvard.edu/abs/2007ApJ...670..668B}
  {670, 668}

\bibitem[\protect\citeauthoryear{{Bogdanov}, {Grindlay}  \&
  {Rybicki}}{{Bogdanov} et~al.}{2008}]{bogdanov08}
{Bogdanov} S.,  {Grindlay} J.~E.,   {Rybicki} G.~B.,  2008, \mn@doi [\apj]
  {10.1086/592341}, \href {http://adsabs.harvard.edu/abs/2008ApJ...689..407B}
  {689, 407}

\bibitem[\protect\citeauthoryear{{Bogdanov}, {van den Berg}, {Heinke}, {Cohn},
  {Lugger}  \& {Grindlay}}{{Bogdanov} et~al.}{2010}]{bogdanov10}
{Bogdanov} S.,  {van den Berg} M.,  {Heinke} C.~O.,  {Cohn} H.~N.,  {Lugger}
  P.~M.,   {Grindlay} J.~E.,  2010, \mn@doi [\apj]
  {10.1088/0004-637X/709/1/241}, \href
  {http://adsabs.harvard.edu/abs/2010ApJ...709..241B} {709, 241}

\bibitem[\protect\citeauthoryear{{Cheng}, {Ho}  \& {Ruderman}}{{Cheng}
  et~al.}{1986}]{cheng86}
{Cheng} K.~S.,  {Ho} C.,   {Ruderman} M.,  1986, \mn@doi [\apj]
  {10.1086/163830}, \href {http://adsabs.harvard.edu/abs/1986ApJ...300..522C}
  {300, 522}

\bibitem[\protect\citeauthoryear{{Deller}, {Verbiest}, {Tingay}  \&
  {Bailes}}{{Deller} et~al.}{2008}]{deller08}
{Deller} A.~T.,  {Verbiest} J.~P.~W.,  {Tingay} S.~J.,   {Bailes} M.,  2008,
  \mn@doi [\apjl] {10.1086/592401}, \href
  {http://adsabs.harvard.edu/abs/2008ApJ...685L..67D} {685, L67}

\bibitem[\protect\citeauthoryear{{Durant}, {Kargaltsev}, {Pavlov}, {Kowalski},
  {Posselt}, {van Kerkwijk}  \& {Kaplan}}{{Durant} et~al.}{2012}]{durant12}
{Durant} M.,  {Kargaltsev} O.,  {Pavlov} G.~G.,  {Kowalski} P.~M.,  {Posselt}
  B.,  {van Kerkwijk} M.~H.,   {Kaplan} D.~L.,  2012, \mn@doi [\apj]
  {10.1088/0004-637X/746/1/6}, \href
  {http://adsabs.harvard.edu/abs/2012ApJ...746....6D} {746, 6}

\bibitem[\protect\citeauthoryear{{Edwards}, {Hobbs}  \& {Manchester}}{{Edwards}
  et~al.}{2006}]{edwards06}
{Edwards} R.~T.,  {Hobbs} G.~B.,   {Manchester} R.~N.,  2006, \mn@doi [\mnras]
  {10.1111/j.1365-2966.2006.10870.x}, \href
  {http://adsabs.harvard.edu/abs/2006MNRAS.372.1549E} {372, 1549}

\bibitem[\protect\citeauthoryear{{Gendreau}, {Arzoumanian}  \&
  {Okajima}}{{Gendreau} et~al.}{2012}]{gendreau12}
{Gendreau} K.~C.,  {Arzoumanian} Z.,   {Okajima} T.,  2012, in Society of
  Photo-Optical Instrumentation Engineers (SPIE) Conference Series. ,
  \mn@doi{10.1117/12.926396}

\bibitem[\protect\citeauthoryear{{Gonzalez} \& {Reisenegger}}{{Gonzalez} \&
  {Reisenegger}}{2010}]{gonzalez10}
{Gonzalez} D.,  {Reisenegger} A.,  2010, \mn@doi [\aap]
  {10.1051/0004-6361/201015084}, \href
  {http://adsabs.harvard.edu/abs/2010A%26A...522A..16G} {522, A16}

\bibitem[\protect\citeauthoryear{{Guillot} \& {Rutledge}}{{Guillot} \&
  {Rutledge}}{2014}]{guillot14}
{Guillot} S.,  {Rutledge} R.~E.,  2014, \mn@doi [\apjl]
  {10.1088/2041-8205/796/1/L3}, \href
  {http://adsabs.harvard.edu/abs/2014ApJ...796L...3G} {796, L3}

\bibitem[\protect\citeauthoryear{{Guillot}, {Rutledge}, {Brown}, {Pavlov}  \&
  {Zavlin}}{{Guillot} et~al.}{2011}]{guillot11b}
{Guillot} S.,  {Rutledge} R.~E.,  {Brown} E.~F.,  {Pavlov} G.~G.,   {Zavlin}
  V.~E.,  2011, \mn@doi [\apj] {10.1088/0004-637X/738/2/129}, \href
  {http://adsabs.harvard.edu/abs/2011ApJ...738..129G} {738, 129}

\bibitem[\protect\citeauthoryear{{Guillot}, {Servillat}, {Webb}  \&
  {Rutledge}}{{Guillot} et~al.}{2013}]{guillot13}
{Guillot} S.,  {Servillat} M.,  {Webb} N.~A.,   {Rutledge} R.~E.,  2013,
  \mn@doi [\apj] {10.1088/0004-637X/772/1/7}, \href
  {http://adsabs.harvard.edu/abs/2013ApJ...772....7G} {772, 7}

\bibitem[\protect\citeauthoryear{{G{\"u}ver}, {{\"O}zel}, {Cabrera-Lavers}  \&
  {Wroblewski}}{{G{\"u}ver} et~al.}{2010}]{guver10a}
{G{\"u}ver} T.,  {{\"O}zel} F.,  {Cabrera-Lavers} A.,   {Wroblewski} P.,  2010,
  \mn@doi [\apj] {10.1088/0004-637X/712/2/964}, \href
  {http://adsabs.harvard.edu/abs/2010ApJ...712..964G} {712, 964}

\bibitem[\protect\citeauthoryear{{Haakonsen}, {Turner}, {Tacik}  \&
  {Rutledge}}{{Haakonsen} et~al.}{2012}]{haakonsen12}
{Haakonsen} C.~B.,  {Turner} M.~L.,  {Tacik} N.~A.,   {Rutledge} R.~E.,  2012,
  \mn@doi [\apj] {10.1088/0004-637X/749/1/52}, \href
  {http://adsabs.harvard.edu/abs/2012ApJ...749...52H} {749, 52}

\bibitem[\protect\citeauthoryear{{Halpern}, {Martin}  \& {Marshall}}{{Halpern}
  et~al.}{1996}]{halpern96b}
{Halpern} J.~P.,  {Martin} C.,   {Marshall} H.~L.,  1996, \mn@doi [\apjl]
  {10.1086/310385}, \href {http://adsabs.harvard.edu/abs/1996ApJ...473L..37H}
  {473, L37}

\bibitem[\protect\citeauthoryear{{Harding}, {Stern}, {Dyks}  \&
  {Frackowiak}}{{Harding} et~al.}{2008}]{harding08}
{Harding} A.~K.,  {Stern} J.~V.,  {Dyks} J.,   {Frackowiak} M.,  2008, \mn@doi
  [\apj] {10.1086/588037}, \href
  {http://adsabs.harvard.edu/abs/2008ApJ...680.1378H} {680, 1378}

\bibitem[\protect\citeauthoryear{{Harrison} et~al.,}{{Harrison}
  et~al.}{2013}]{harrison13}
{Harrison} F.~A.,  et~al., 2013, \mn@doi [\apj] {10.1088/0004-637X/770/2/103},
  \href {http://adsabs.harvard.edu/abs/2013ApJ...770..103H} {770, 103}

\bibitem[\protect\citeauthoryear{{Heinke}}{{Heinke}}{2013}]{heinke13}
{Heinke} C.~O.,  2013, \mn@doi [Journal of Physics Conference Series]
  {10.1088/1742-6596/432/1/012001}, \href
  {http://adsabs.harvard.edu/abs/2013JPhCS.432a2001H} {432, 012001}

\bibitem[\protect\citeauthoryear{{Heinke}, {Rybicki}, {Narayan}  \&
  {Grindlay}}{{Heinke} et~al.}{2006}]{heinke06a}
{Heinke} C.~O.,  {Rybicki} G.~B.,  {Narayan} R.,   {Grindlay} J.~E.,  2006,
  \mn@doi [\apj] {10.1086/503701}, \href
  {http://adsabs.harvard.edu/abs/2006ApJ...644.1090H} {644, 1090}

\bibitem[\protect\citeauthoryear{{Heinke} et~al.,}{{Heinke}
  et~al.}{2014}]{heinke14}
{Heinke} C.~O.,  et~al., 2014, \mn@doi [\mnras] {10.1093/mnras/stu1449}, \href
  {http://adsabs.harvard.edu/abs/2014MNRAS.444..443H} {444, 443}

\bibitem[\protect\citeauthoryear{{Johnston} et~al.,}{{Johnston}
  et~al.}{1993}]{johnston93}
{Johnston} S.,  et~al., 1993, \mn@doi [\nat] {10.1038/361613a0}, \href
  {http://adsabs.harvard.edu/abs/1993Natur.361..613J} {361, 613}

\bibitem[\protect\citeauthoryear{{Kargaltsev}, {Pavlov}  \&
  {Romani}}{{Kargaltsev} et~al.}{2004}]{kargaltsev04}
{Kargaltsev} O.,  {Pavlov} G.~G.,   {Romani} R.~W.,  2004, \mn@doi [\apj]
  {10.1086/380993}, \href {http://adsabs.harvard.edu/abs/2004ApJ...602..327K}
  {602, 327}

\bibitem[\protect\citeauthoryear{{Kaspi}, {Roberts}  \& {Harding}}{{Kaspi}
  et~al.}{2006}]{kaspi06}
{Kaspi} V.~M.,  {Roberts} M.~S.~E.,   {Harding} A.~K.,  2006, {Isolated neutron
  stars}.
pp 279--339, \mn@doi{10.2277/0521826594}

\bibitem[\protect\citeauthoryear{{Krimm} et~al.,}{{Krimm}
  et~al.}{2013}]{krimm13}
{Krimm} H.~A.,  et~al., 2013, \mn@doi [\apjs] {10.1088/0067-0049/209/1/14},
  \href {http://adsabs.harvard.edu/abs/2013ApJS..209...14K} {209, 14}

\bibitem[\protect\citeauthoryear{{Lattimer} \& {Prakash}}{{Lattimer} \&
  {Prakash}}{2007}]{lattimer07}
{Lattimer} J.~M.,  {Prakash} M.,  2007, \mn@doi [\physrep]
  {10.1016/j.physrep.2007.02.003}, \href
  {http://adsabs.harvard.edu/abs/2007PhR...442..109L} {442, 109}

\bibitem[\protect\citeauthoryear{{Lyutikov}}{{Lyutikov}}{2013}]{lyutikov13}
{Lyutikov} M.,  2013, \mn@doi [\mnras] {10.1093/mnras/stt351}, \href
  {http://adsabs.harvard.edu/abs/2013MNRAS.431.2580L} {431, 2580}

\bibitem[\protect\citeauthoryear{{Madsen} et~al.,}{{Madsen}
  et~al.}{2015}]{madsen15}
{Madsen} K.~K.,  et~al., 2015, \mn@doi [\apjs] {10.1088/0067-0049/220/1/8},
  \href {http://adsabs.harvard.edu/abs/2015ApJS..220....8M} {220, 8}

\bibitem[\protect\citeauthoryear{{Marelli}, {De Luca}  \& {Caraveo}}{{Marelli}
  et~al.}{2011}]{marelli11}
{Marelli} M.,  {De Luca} A.,   {Caraveo} P.~A.,  2011, \mn@doi [\apj]
  {10.1088/0004-637X/733/2/82}, \href
  {http://adsabs.harvard.edu/abs/2011ApJ...733...82M} {733, 82}

\bibitem[\protect\citeauthoryear{{Miller}}{{Miller}}{2013}]{miller13}
{Miller} M.~C.,  2013, preprint, \href
  {http://adsabs.harvard.edu/abs/2013arXiv1312.0029M} {} (\mn@eprint {arXiv}
  {1312.0029})

\bibitem[\protect\citeauthoryear{{Ng}, {Takata}, {Leung}, {Cheng}  \&
  {Philippopoulos}}{{Ng} et~al.}{2014}]{ng14}
{Ng} C.-Y.,  {Takata} J.,  {Leung} G.~C.~K.,  {Cheng} K.~S.,   {Philippopoulos}
  P.,  2014, \mn@doi [\apj] {10.1088/0004-637X/787/2/167}, \href
  {http://adsabs.harvard.edu/abs/2014ApJ...787..167N} {787, 167}

\bibitem[\protect\citeauthoryear{{{\"O}zel} \& {Psaltis}}{{{\"O}zel} \&
  {Psaltis}}{2009}]{ozel09b}
{{\"O}zel} F.,  {Psaltis} D.,  2009, \mn@doi [\prd]
  {10.1103/PhysRevD.80.103003}, \href
  {http://adsabs.harvard.edu/abs/2009PhRvD..80j3003O} {80, 103003}

\bibitem[\protect\citeauthoryear{{{\"O}zel}, {Psaltis}, {G{\"u}ver}, {Baym},
  {Heinke}  \& {Guillot}}{{{\"O}zel} et~al.}{2016}]{ozel16}
{{\"O}zel} F.,  {Psaltis} D.,  {G{\"u}ver} T.,  {Baym} G.,  {Heinke} C.,
  {Guillot} S.,  2016, \mn@doi [\apj] {10.3847/0004-637X/820/1/28}, \href
  {http://adsabs.harvard.edu/abs/2016ApJ...820...28O} {820, 28}

\bibitem[\protect\citeauthoryear{{Reardon} et~al.,}{{Reardon}
  et~al.}{2016}]{reardon16}
{Reardon} D.~J.,  et~al., 2016, \mn@doi [\mnras] {10.1093/mnras/stv2395}, \href
  {http://adsabs.harvard.edu/abs/2016MNRAS.455.1751R} {455, 1751}

\bibitem[\protect\citeauthoryear{{Romani}}{{Romani}}{1996}]{romani96}
{Romani} R.~W.,  1996, \mn@doi [\apj] {10.1086/177878}, \href
  {http://adsabs.harvard.edu/abs/1996ApJ...470..469R} {470, 469}

\bibitem[\protect\citeauthoryear{{Suleimanov}, {Poutanen}, {Revnivtsev}  \&
  {Werner}}{{Suleimanov} et~al.}{2011}]{suleimanov11b}
{Suleimanov} V.,  {Poutanen} J.,  {Revnivtsev} M.,   {Werner} K.,  2011,
  \mn@doi [\apj] {10.1088/0004-637X/742/2/122}, \href
  {http://adsabs.harvard.edu/abs/2011ApJ...742..122S} {742, 122}

\bibitem[\protect\citeauthoryear{{Verbiest} et~al.,}{{Verbiest}
  et~al.}{2008}]{verbiest08}
{Verbiest} J.~P.~W.,  et~al., 2008, \mn@doi [\apj] {10.1086/529576}, \href
  {http://adsabs.harvard.edu/abs/2008ApJ...679..675V} {679, 675}

\bibitem[\protect\citeauthoryear{{Webb} \& {Barret}}{{Webb} \&
  {Barret}}{2007}]{webb07}
{Webb} N.~A.,  {Barret} D.,  2007, \mn@doi [\apj] {10.1086/522877}, \href
  {http://adsabs.harvard.edu/abs/2007ApJ...671..727W} {671, 727}

\bibitem[\protect\citeauthoryear{{Wijnands} \& {van der Klis}}{{Wijnands} \&
  {van der Klis}}{1998}]{wijnands98}
{Wijnands} R.,  {van der Klis} M.,  1998, \mn@doi [\nat] {10.1038/28557}, \href
  {http://adsabs.harvard.edu/abs/1998Natur.394..344W} {394, 344}

\bibitem[\protect\citeauthoryear{{Wilms}, {Allen}  \& {McCray}}{{Wilms}
  et~al.}{2000}]{wilms00}
{Wilms} J.,  {Allen} A.,   {McCray} R.,  2000, \mn@doi [\apj] {10.1086/317016},
  \href {http://adsabs.harvard.edu/abs/2000ApJ...542..914W} {542, 914}

\bibitem[\protect\citeauthoryear{{Winkler} et~al.,}{{Winkler}
  et~al.}{2003}]{winkler03}
{Winkler} C.,  et~al., 2003, \mn@doi [\aap] {10.1051/0004-6361:20031288}, \href
  {http://adsabs.harvard.edu/abs/2003A%26A...411L...1W} {411, L1}

\bibitem[\protect\citeauthoryear{{Zavlin}}{{Zavlin}}{2006}]{zavlin06}
{Zavlin} V.~E.,  2006, \mn@doi [\apj] {10.1086/449308}, \href
  {http://adsabs.harvard.edu/abs/2006ApJ...638..951Z} {638, 951}

\bibitem[\protect\citeauthoryear{{Zavlin}, {Pavlov}  \& {Shibanov}}{{Zavlin}
  et~al.}{1996}]{zavlin96}
{Zavlin} V.~E.,  {Pavlov} G.~G.,   {Shibanov} Y.~A.,  1996, \aap, \href
  {http://adsabs.harvard.edu/abs/1996A%26A...315..141Z} {315, 141}

\bibitem[\protect\citeauthoryear{{Zavlin}, {Pavlov}, {Sanwal}, {Manchester},
  {Tr{\"u}mper}, {Halpern}  \& {Becker}}{{Zavlin} et~al.}{2002}]{zavlin02b}
{Zavlin} V.~E.,  {Pavlov} G.~G.,  {Sanwal} D.,  {Manchester} R.~N.,
  {Tr{\"u}mper} J.,  {Halpern} J.~P.,   {Becker} W.,  2002, \mn@doi [\apj]
  {10.1086/339351}, \href {http://adsabs.harvard.edu/abs/2002ApJ...569..894Z}
  {569, 894}

\bibitem[\protect\citeauthoryear{{de Jager}, {Raubenheimer}  \&
  {Swanepoel}}{{de Jager} et~al.}{1989}]{dejager89}
{de Jager} O.~C.,  {Raubenheimer} B.~C.,   {Swanepoel} J.~W.~H.,  1989, \aap,
  \href {http://adsabs.harvard.edu/abs/1989A%26A...221..180D} {221, 180}

\makeatother
\end{thebibliography}

\appendix

\section{PSR J0437--4715 Ephemeris}
\label{sec:appendix}

We provide in Table~\ref{tab:ephem} of this appendix the parameters of
the ephemeris obtained from data collected at the Molonglo observatory
between March 2005 and February 2011.

\begin{table} 
 \centering
 \caption{Molonglo ephemeris of PSR~J0437--4715. 1 $\sigma$
   uncertainties on the last digit of each parameters are represented
   in parantheses.}
% \resizebox{\columnwidth}{!}{%
   \begin{tabular}{lc}
     \hline
     Parameter & Value \\
     \hline
      Right ascension $\alpha$ (J2000)      &  04 37 15.8961749(7)  \\
      Declination $\delta$ (J2000)          & -47 15 09.11071(1)   \\
      Proper motion in $\alpha$ (mas/yr)    & 121.428(4) \\
      Proper motion in $\delta$ (mas/yr)    & -71.473(5) \\
      Annual parallax (mas)                 & 6.28 (11)\\
      \hline
      Pulse period (ms)                     & 5.757451936712643(3) \\
      Pulse period derivative ($10^{-20}$)   & 5.72918(1) \\
      Orbital period (days)                 & 5.7410480(9) \\
      Orbital period derivative ($10^{-12}$) & 3.75(4) \\ 
      Epoch of periastron passage (MJD)     & 54530.1724(3) \\
      Projected semi-major axis (s)         & 3.36671468(5) \\
      Orbital eccentricity ($10^{-5}$)      & 1.917974290(1) \\
      Longitude of periastron (\deg)        & 1.37(2) \\
      Periastron advance (\deg/yr)          & 0.022(4) \\
      Longitude of ascension (\deg)         & 207(1) \\
      Orbital inclination (\deg)            & 137.55(5) \\
      \hline
      Reference epoch (MJD)                 & 54500 \\
      MJD Range                             & 53431.26--55619.19 \\
      \hline
     \label{tab:ephem}
   \end{tabular}  % }
\end{table} 

\end{document}